\documentclass[aps,prl,twocolumn,superscriptaddress]{revtex4-1}
\usepackage{amssymb, amsmath}
\usepackage{graphicx,onlyamsmath}

\usepackage{natbib}
\usepackage{xcolor}
\usepackage[normalem]{ulem}  

\begin{document}

\title{Massless Dirac fermions in III-V semiconductor quantum wells}

\author{S.~S.~Krishtopenko}
\thanks{These two authors contributed equally.}
\affiliation{Laboratoire Charles Coulomb, Universit\'{e} de Montpellier, Centre National de la Recherche Scientifique, 34095 Montpellier, France.}
\affiliation{Institute for Physics of Microstructures RAS, GSP-105, Nizhni Novgorod 603950, Russia.}

\author{W.~Desrat}
\thanks{These two authors contributed equally.}
\affiliation{Laboratoire Charles Coulomb, Universit\'{e} de Montpellier, Centre National de la Recherche Scientifique, 34095 Montpellier, France.}

\author{K.~E.~Spirin}
\affiliation{Institute for Physics of Microstructures RAS, GSP-105, Nizhni Novgorod 603950, Russia.}

\author{C.~Consejo}
\affiliation{Laboratoire Charles Coulomb, Universit\'{e} de Montpellier, Centre National de la Recherche Scientifique, 34095 Montpellier, France.}

\author{S.~Ruffenach}
\affiliation{Laboratoire Charles Coulomb, Universit\'{e} de Montpellier, Centre National de la Recherche Scientifique, 34095 Montpellier, France.}

\author{F.~Gonzalez-Posada}
\affiliation{Institut d'Electronique et des Syst\`emes, Universit\'{e} de Montpellier, Centre National de la Recherche Scientifique, 34000 Montpellier, France.}

\author{B.~Jouault}
\affiliation{Laboratoire Charles Coulomb, Universit\'{e} de Montpellier, Centre National de la Recherche Scientifique, 34095 Montpellier, France.}

\author{W.~Knap}
\affiliation{Laboratoire Charles Coulomb, Universit\'{e} de Montpellier, Centre National de la Recherche Scientifique, 34095 Montpellier, France.}

\author{K.~V.~Maremyanin}
\affiliation{Institute for Physics of Microstructures RAS, GSP-105, Nizhni Novgorod 603950, Russia.}

\author{V.~I.~Gavrilenko}
\affiliation{Institute for Physics of Microstructures RAS, GSP-105, Nizhni Novgorod 603950, Russia.}

\author{G.~Boissier}
\affiliation{Institut d'Electronique et des Syst\`emes, Universit\'{e} de Montpellier, Centre National de la Recherche Scientifique, 34000 Montpellier, France.}

\author{J.~Torres}
\affiliation{Institut d'Electronique et des Syst\`emes, Universit\'{e} de Montpellier, Centre National de la Recherche Scientifique, 34000 Montpellier, France.}

\author{M.~Zaknoune}
\affiliation{Institut d'Electronique, de Micro\'electronique et de Nanotechnologie, Lille University, Avenue Poincar\'e, B.P. 60069, F-59652 Villeneuve d'Ascq, France}

\author{E.~Tourni\'{e}}
\affiliation{Institut d'Electronique et des Syst\`emes, Universit\'{e} de Montpellier, Centre National de la Recherche Scientifique, 34000 Montpellier, France.}

\author{F.~Teppe}
\email[]{frederic.teppe@umontpellier.fr}
\affiliation{Laboratoire Charles Coulomb, Universit\'{e} de Montpellier, Centre National de la Recherche Scientifique, 34095 Montpellier, France.}
\date{\today}

\begin{abstract}
We report on the clear evidence of massless Dirac fermions in two-dimensional system based on III-V semiconductors. Using a gated Hall bar made on a three-layer InAs/GaSb/InAs quantum well, we restore the Landau levels fan chart by magnetotransport and unequivocally demonstrate a gapless state in our sample. Measurements of cyclotron resonance at different electron concentrations directly indicate a linear band crossing at the $\Gamma$ point of Brillouin zone. Analysis of experimental data within analytical Dirac-like Hamiltonian allows us not only determing velocity $v_F=1.8\cdot10^5$~m/s of massless Dirac fermions but also demonstrating significant non-linear dispersion at high energies.
\end{abstract}

\pacs{73.21.Fg, 73.43.Lp, 73.61.Ey, 75.30.Ds, 75.70.Tj, 76.60.-k} 
\keywords{}
\maketitle

Since relativistic Dirac-like character of charge carriers was demonstrated in monolayer graphene~\cite{Q1}, two-dimensional (2D) massless Dirac fermions (DFs) are intensively studied in condensed matter physics. There are several systems~\cite{Q2} from graphene-like 2D materials (silicene, germanene, etc.) or high-temperature $d$-wave superconductors to the surfaces of three-dimensional (3D) topological insulators, in which the presence of 2D massless DFs was revealed. Their universal features, such as suppressed backscattering~\cite{Q3}, Klein tunneling~\cite{Q1c}, giant magnetoresistance~\cite{Q1d}, or their specific response to impurities and magnetic field~\cite{Q1b} hold great promises for new nano-scale electronic devices.

Among quantum well (QW) systems, a single-valley spin-degenerate Dirac cone at the $\Gamma$ point of Brillouin zone was theoretically predicted~\cite{Q4,Q5} and experimentally observed~\cite{Q6a,Q6b,Q6c,Q6d} in HgTe/CdTe QWs. At a critical width, the band gap in these QWs vanishes and the band structure changes from trivial to inverted. The key advantage of QWs over other systems is based on the ability to adjust DFs velocity by adjusting the strain and thickness of the layers. It allows varying the ratio between kinetic energy and Coulomb interaction, which results in a rich variety of phenomena involving massless DFs~\cite{Q13}. However, the massless DFs in HgTe QWs appear only at a fixed temperature, since the temperature changes open a band gap resulting in a non-zero rest-mass of the particles~\cite{Q7,Q8,Q9,Q10,Q11}.

Searching for 2D massless DFs in other QWs, some authors considered theoretically very thin (few atomic layers) conventional III-V semiconductor heterostructures, like GaN/InN/GaN~\cite{Q14} and GaAs/Ge/GaAs QWs~\cite{Q15}. Depending on the number of atomic layers in these QWs, the band structure can be trivial, inverted, or gapless, just like in HgTe QWs~\cite{Q4,Q5}. Although considerable progress in the fabrication of GaN/InN/GaN and GaAs/Ge/GaAs structures was obtained, there are still no experimental results confirming the presence of massless DFs in these structures.

Alternative III-V semiconductor QWs, in which massless DFs have been theoretically predicted, are symmetric three-layer InAs/Ga$_{x}$In$_{1-x}$Sb/InAs QWs confined between wide-gap AlSb barriers~\cite{Q16}. Depending on their layer thicknesses, these QWs host trivial, quantum spin Hall insulator and gapless states. However, in contrast to the HgTe QWs, the three-layer QWs have a temperature-insensitive band-gap, as it has been recently shown by terahertz spectroscopy~\cite{Q17}. Another difference of massless DFs in InAs/Ga$_{x}$In$_{1-x}$Sb/InAs QWs is recently predicted~\cite{Q16} large tunability of quasiparticle's velocity, which can be varied from $1\cdot10^5$~m/s to $7\cdot10^5$~m/s depending on $x$ and the layer thicknesses. The latter offers the possibility not only to tune the electronic properties~\cite{Q3,Q1b,Q1c} of the DFs but also to achieve specific non-trivial states induced by electron-electron interaction~\cite{Q19,Q20,Q21,Q22}.

In this work, we report striking evidence of the presence of massless Dirac fermions in InAs/GaSb/InAs QWs embedded between AlSb barriers. Measuring magnetoresistance of a gated Hall bar, we restore the Landau level (LL) fan chart in our sample, as firstly performed by B\"{u}ttner et al.~\cite{Q6a} in HgTe QWs. Our experimental data clearly evidence a gapless state. We also measure cyclotron resonance (CR) at different electron concentrations varied by bipolar persistent photoconductivity (PPC) inherent to InAs-based QWs~\cite{Q23,Q24,Q25,Q26,Q27,Q28,Q18,Q28a}. The latter acts as an optical gating and allows changing the electron concentration in the QW by several times. By analyzing the dependence of the cyclotron mass as a function of the concentration, the massless DF velocity is deduced. To analyze these data, we use both realistic band structure calculations based on an eight-band Kane model~\cite{Q16} and analytical approach involving a simplified Dirac-like Hamiltonian~\cite{Q5}.

\begin{figure}
\includegraphics [width=0.95\columnwidth, keepaspectratio] {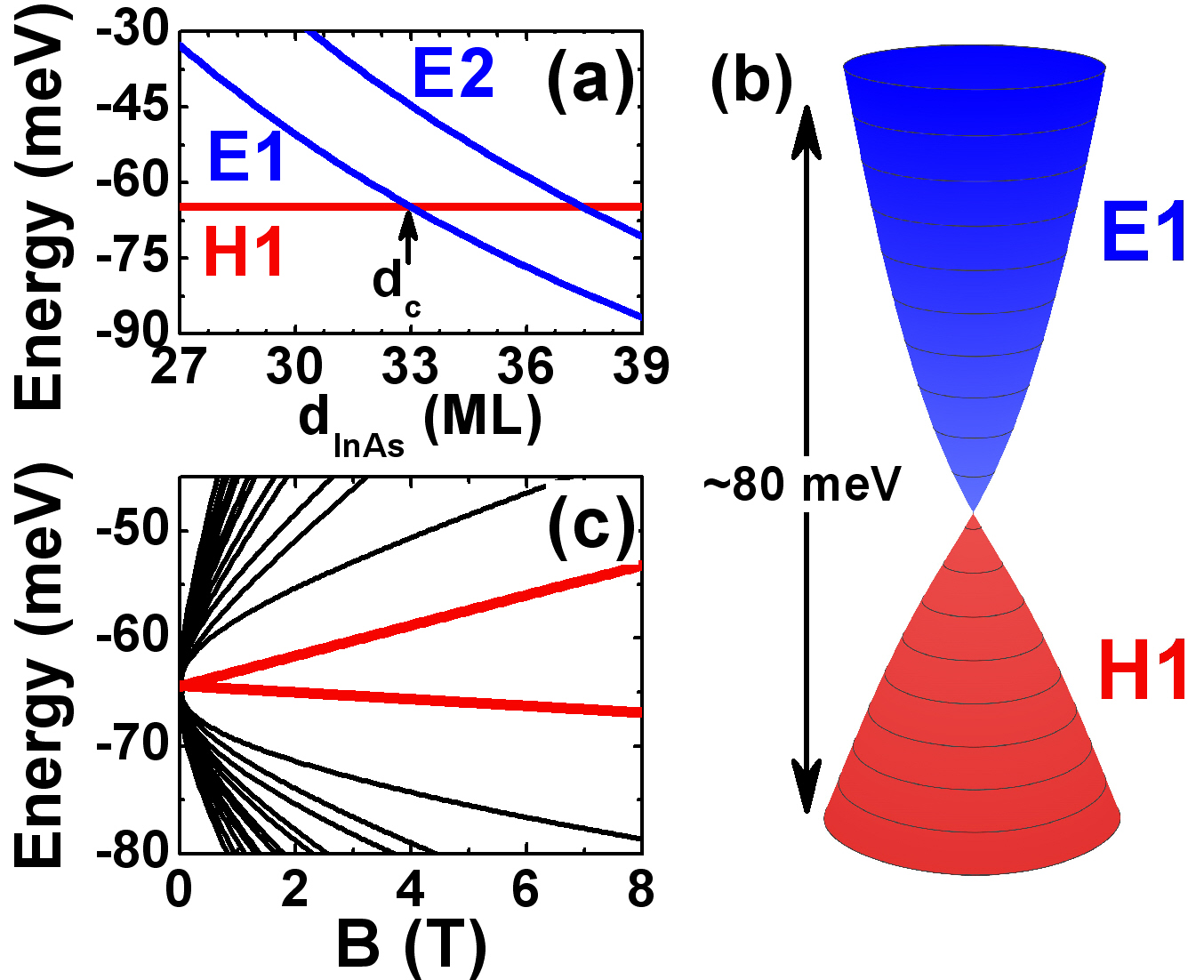} 
\caption{\label{Fig:1} (color online) (a) Subband energies at zero quasimomentum $\mathbf{k}$ in symmetric InAs/GaSb/InAs QWs as a function of InAs layer thickness $d_{\mathrm{InAs}}$. Blue and red curves correspond to the electron-like and hole-like states, respectively. The thickness of the GaSb layer $d_{\mathrm{GaSb}}$ equals to 12 monolayers (ML). Here, 1 ML corresponds to half of the lattice constant of the bulk material. The zero energy is referenced to the valence-band edge of bulk GaSb. (b) A 3D plot of the Dirac cone describing the energy spectrum at $\mathbf{k}<0.2$~nm$^{-1}$ for $d_{\mathrm{InAs}}=33$~ML and $d_{\mathrm{GaSb}}=12$~ML. (c) LL fan chart for the gapless state. A pair of zero-mode LLs is presented by the red curves. All the calculations are based on an eight-band Kane model~\cite{Q16}.}
\end{figure}

As mentioned above, the band structure of three-layer InAs/GaSb/InAs QWs, related to the mutual position of electron-like and hole-like subbands, strongly depends on the layer thicknesses~\cite{Q16}. When the InAs and GaSb layers are both thin, the first electron-like (\emph{E}1) and hole-like (\emph{H}1) subbands correspond to the conduction and valence bands respectively and the QW has a trivial band ordering. For thicker layers, the \emph{E}1 subband drops below the \emph{H}1 subband, as shown in Fig.~\ref{Fig:1}a, and the system has an inverted band ordering. In this case, the conduction and valence bands are represented by the hole-like and electron-like levels, respectively.


One can use a simplified Dirac-like Hamiltonian~\cite{Q5} to describe the electronic states when the energy difference between \emph{E}1 and \emph{H}1 subbands is small. Within the representation defined by the basis $|$\emph{E}1,+$\rangle$, $|$\emph{H}1,+$\rangle$, $|$\emph{E}1,-$\rangle$, $|$\emph{H}1,-$\rangle$, it has the form:
\begin{equation*}
H(\mathbf{k})=\begin{pmatrix}
H_{\mathrm{D}}(\mathbf{k}) & 0 \\ 0 & H_{\mathrm{D}}^{*}(-\mathbf{k})\end{pmatrix},~H_{\mathrm{D}}(\mathbf{k})=\epsilon(\textbf{k})+\sum_{i=1}^3d_i(\textbf{k})\sigma_i,
\end{equation*}
where asterisk stands for complex conjugation, $\mathbf{k}=(k_x,k_y)$ is the momentum in the QW plane, $\sigma_i$ are the Pauli matrices, $\epsilon_{\mathbf{k}}=C-D(k_x^2+k_y^2)$, $d_1(\mathbf{k})=-Ak_x$, $d_2(\mathbf{k})=-Ak_y$, and $d_3(\mathbf{k})=M-B(k_x^2+k_y^2)$. The structure parameters $A$, $B$, $C$, $D$, $M$ depend on the layer thicknesses. The mass parameter $M$ is positive for trivial band ordering and negative for inverted band structure. If we only keep the terms up to linear order in $k$ for each spin, then $H_{\mathrm{D}}(\mathbf{k})$ and $H_{\mathrm{D}}^{*}(-\mathbf{k})$ at $M=0$ correspond to massless Dirac Hamiltonians. We note that the latter is valid if the InAs/GaSb/InAs QW has an inversion symmetry in the growth direction~\cite{Q29}. In this case, the \emph{E}1 and \emph{H}1 subbands cross at the $\Gamma$ point, and their energy dispersion calculated from an eight-band Kane model is found to linearly depend on the quasimomentum at small~$\mathbf{k}$, as shown in Fig.~\ref{Fig:1}b.

Besides the linear terms, $H_{\mathrm{D}}(\mathbf{k})$ also contains quadratic terms, which cannot be neglected even at the energies close to the band crossing point. Moreover, they result in relevant difference between conventional massless DFs in graphene~\cite{Q1,Q3,Q1b,Q1c} and the ones in symmetric InAs/GaSb/InAs QWs (and in HgTe QWs~\cite{Q6a,Q6b,Q6c} as well). LLs in graphene are characterized by both a square-root dependence of their energies on the magnetic field and presence of so-called \emph{zero-energy} LLs independent of the field. Note that all LLs in graphene have a spin-degeneracy (for simplicity, we consider the electrons in one valley and neglect the small Zeeman effect). This case is described by the linear terms in $H_{\mathrm{D}}(\mathbf{k})$ and $H_{\mathrm{D}}^{*}(-\mathbf{k})$ at $M=0$~\cite{SM}.

The parabolic terms remove the spin degeneracy of all LLs~\cite{SM} and, particularly, transform the spin-degenerate \emph{zero-energy} LL into a pair of spin-polarized \emph{zero-mode} LLs~\cite{Q30}, as shown in Fig.~\ref{Fig:1}c. The electron-like zero-mode LL splits from the edge of the \emph{E}1 subband and tends toward high energy with increasing magnetic field. In contrast, the second level, which decreases with magnetic field, has a hole-like character and arises from the \emph{H}1 subband. Therefore, the crossing of the zero-mode LLs at finite value of magnetic field occurs in the inverted region, $M<0$ and is absent for $M>0$~\cite{Q10}. A crossing of the zero-mode LLs at zero magnetic field gives a direct indication for the massless DFs in the QW.

\begin{figure}
\includegraphics [width=0.95\columnwidth, keepaspectratio] {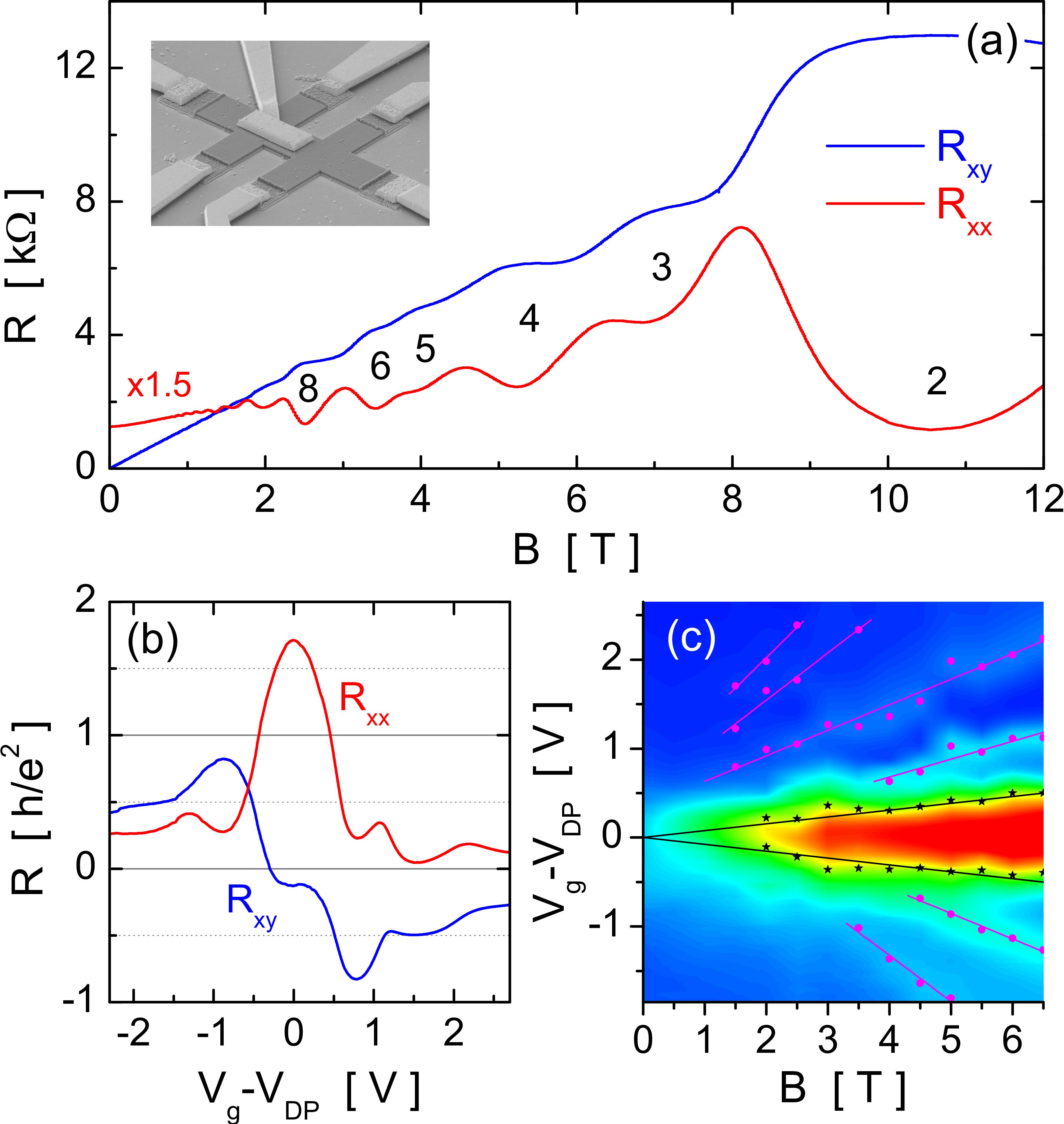} 
\caption{\label{Fig:2} (color online) (a) Longitudinal and Hall resistances as a function of magnetic field at $T=1.7$~K and $V_g=0$~V. Electron microscopy image of the Hall bar is shown in the inset. (b) $R_{xx}$ and $R_{xy}$ versus $V_g$ at $T=1.7$~K and $B=6.5$~T. (c) Color map of the longitudinal resistance as a function of $V_g$ and $B$. Black stars indicate $h/e^2$ values and red dots the peak values of $R_{xx}$. }
\end{figure}

The sample studied in this work was grown by molecular beam epitaxy (MBE) on a semi-insulating (001) GaAs substrate with a relaxed GaSb buffer. In order to get the gapless state, the thicknesses of InAs and GaSb layers were 33 ML and 14 ML, respectively (see Fig.~\ref{Fig:1}). After the growth, a 50 $\mu$m wide gated Hall bar was fabricated by using a single-mesa process. All details are provided in the Supplemental Materials~\cite{SM}.

First, we investigate magnetotransport in our sample. Figure~\ref{Fig:2}a presents the magnetic field dependence of the longitudinal and Hall resistances at $T=1.7$~K. The Hall density extracted from the measurements is equal to $n_S=5.1\cdot10^{11}$~cm$^{-2}$ with electron carriers at zero gate voltage $V_g$ and a mobility of $\mu_e=1.65\cdot10^4$~cm$^2$V$^{-1}$s$^{-1}$. The Hall resistance shows well-defined plateaus as a function of magnetic field $B$ at both even and odd multiples of $h/e^2$ associated to minima in the Shubnikov-de Haas (ShdH) oscillations, proving a 2D character of charge carriers in our structure. To vary the carrier density in the QW, we apply a gate voltage $V_g$ to the top gate. Typical gate voltage dependencies of $R_{xx}$ and $R_{xy}$ at $B=6.5$~T are shown in Fig.~\ref{Fig:2}b (the low-field data are provided in the Supplemental Materials~\cite{SM}).


\begin{figure}
\includegraphics [width=1.0\columnwidth, keepaspectratio] {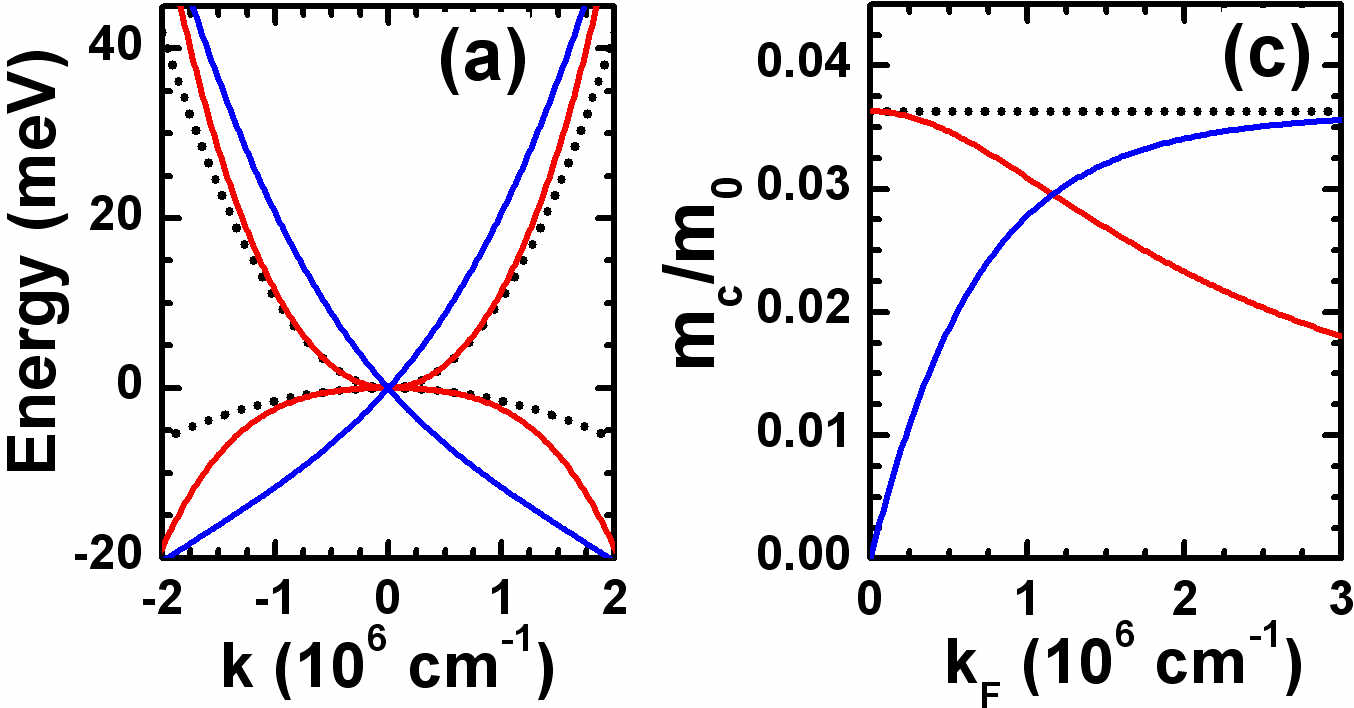} 
\caption{\label{Fig:3new} Band dispersion (a) and cyclotron mass $m_c$ in the conduction band as a function of Fermi momentum, $k_F=\sqrt{2\pi n_S}$ for the gapless states with linear (blue curves) and parabolic band touching at the $\Gamma$ point. Two cases for parabolic touching, when conduction and valence band has the same ($\mathcal{A}=0$) and opposite ($\mathcal{A}\neq 0$) parity~\cite{SM}, are represented by dotted black and solid red curves, respectively. The following band parameters were used in the calculation: $C=0$~meV, $A=150$~meV$\cdot$nm, $\mathcal{A}=35$~meV$\cdot$nm$^3$, $B=-600$~meV$\cdot$nm$^2$ and $D=-450$~meV$\cdot$nm$^2$.}
\end{figure}

The longitudinal resistance shows clear oscillations on each side as a function of $V_g$ of the central peak occurring at $V_{DP}\simeq-4$~V when the Hall resistance presents $\pm0.5\,h/e^2$ plateaus and a sign reversal. These features demonstrate a changing of carrier concentration at different $V_g$ and the inversion of the type of carriers from electrons to holes at large negative gate voltage. Furthermore, an insulating behavior is observed at around $V_g=V_{DP}$ with $R_{xx}>h/e^2$ and small values of $R_{xy}$. The evolution of this insulating state is plotted in a 2D color map of $R_{xx}$ as a function of $V_g$ and $B$ (Fig.~\ref{Fig:2}c). It is evident that the size of the insulating region (red area) increases with $B$. The linear extrapolation down to zero field of the $R_{xx}=h/e^2$ points (black stars), corresponding to the position of the zero-mode LLs~\cite{Q6a,Q10} and delimiting the high resistance region for $B>2$~T, demonstrates that the insulating state vanishes at $B=0$~T. This is also confirmed by the analysis of $d\sigma_{xy}/dV_g$ provided in the Supplemental Materials~\cite{SM}. Thus, the crossing of the zero-mode LLs at $B=0$~T indeed confirms the gapless state in our sample. The traces of higher LLs of electrons and holes are also seen for $V_g-V_{DP}$ higher than $0.5$~V and lower than $-0.5$~V, respectively.

The hallmark of 2D massless DFs is a specific sequence of quantum Hall plateaux observed at odd multiples of $g_ve^2/h$, where is $g_v$ valley degeneracy factor~\cite{Q3}. In HgTe QWs ($g_v=1$), the odd-integer quantum Hall sequence is much less pronounced~\cite{Q6a} than in graphene~\cite{Q1} ($g_v=2$) and observed at small (less than 1~T) magnetic fields only. The latter is caused by prominent contribution of the parabolic terms in $H_{\mathrm{D}}(\mathbf{k})$, which remove the spin degeneracy of all LLs already at moderate fields as discussed above. In our sample, both \emph{odd} and \emph{even} plateaux are observed (see Fig.~\ref{Fig:2}), and the quantum Hall effect looks like the one in conventional 2D electron gas~\cite{Q35}. This may be interpreted in two different ways.

First, large contribution of terms proportional to $B$ and $D$ in $H_{\mathrm{D}}(\mathbf{k})$ results in significant spin splitting of LLs making impossible observation of the odd-integer sequence. Second, our sample may host the gapless state with parabolic band touching. As it is shown in the Supplemental Materials~\cite{SM}, band dispersion  in $k$ for the latter case up to the third order has the form:
\begin{equation}
\label{eq:2new}
E_{k^2}(k)=C-Dk^2\pm k^2\sqrt{B^2+\mathcal{A}^2k^2},
\end{equation}
where "$+$" and "$-$" represent to the conduction and valence bands, respectively. Here, $\mathcal{A}\neq0$ corresponds to the opposite parity of conduction and valence band, while for the same parity, one should set $\mathcal{A}=0$~\cite{SM}. Note that the linear dispersion is described by $E_{k}(k)=C-Dk^2\pm k\sqrt{A^2+B^2k^2}$. The band dispersions at specific parameters are shown in Fig.~\ref{Fig:3new}a.

An efficient way to discriminate these two gapless states is the measurement of quasiclassical CR at different Fermi level positions. Applying a quasiclassical quantization rule to $E_{k}(k)$ and $E_{k^2}(k)$, the cyclotron mass $m_c$ in the conduction band as a function of Fermi momentum $k_F=\sqrt{2\pi n_S}$ has the forms
\begin{equation}
\label{eq:3}
m^{(k)}_{c}(k_F)=\dfrac{\hbar^2k_F\sqrt{A^2+B^2k_F^2}}{A^2+2B^2k_F^2-2Dk_F\sqrt{A^2+B^2k_F^2}}
\end{equation}
and
\begin{equation}
\label{eq:4}
m^{(k^2)}_{c}(k_F)=\dfrac{\hbar^2\sqrt{B^2+\mathcal{A}^2k_F^2}}{2B^2+3\mathcal{A}^2k_F^2-2D\sqrt{B^2+\mathcal{A}^2k_F^2}}
\end{equation}
for the linear and parabolic case, respectively. As it is seen from Fig.~\ref{Fig:3new}b, the linear and parabolic gapless states have different behavior of cyclotron mass as a function of $k_F$.

\begin{figure}
\includegraphics [width=1.0\columnwidth, keepaspectratio] {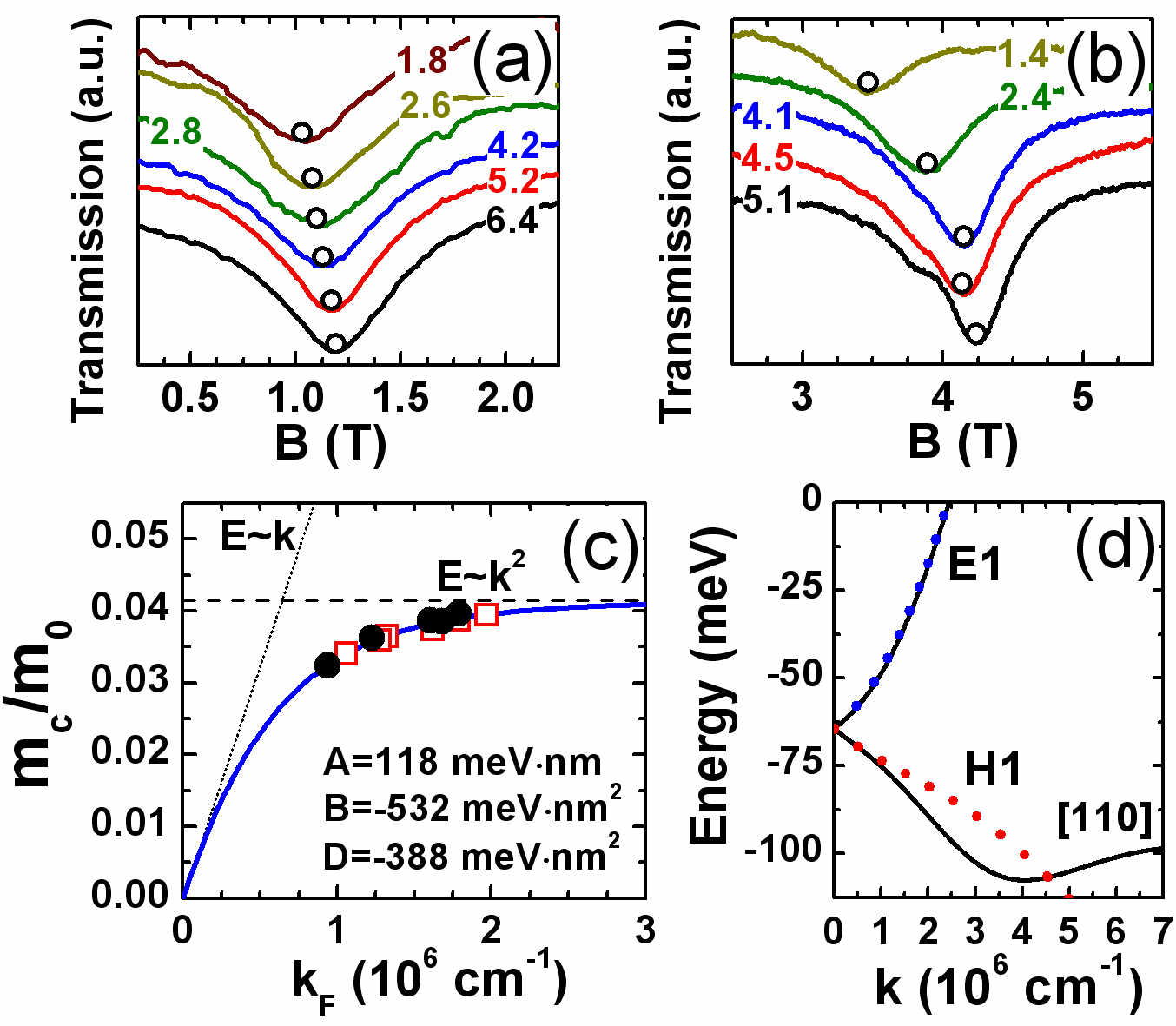} 
\caption{\label{Fig:4} (color online) (a,b) CR spectra at $T=4.2$~K of electrons measured with BWO at 845 GHz (a) and QCL at the 3 THz (b) for different values of electron concentration. The symbols mark positions of CR lines. The numbers over the curves show $n_S$ in $10^{11}$~cm$^{-2}$. (c) Cyclotron mass $m_c$ as a function of Fermi momentum, $k_F=\sqrt{2\pi n_S}$. The open and solid symbols correspond to the values obtained by BWO and QCL, respectively. The bold blue curve is the fitting to Eq.~(\ref{eq:3}). (d) Comparison of band dispersion calculated within the eight-band Kane model with the one described by a simplified Dirac-like Hamiltonian with the parameters shown in the panel (c). Parameter $C$ is chosen to have a coincidence of the band crossing point in two models.}
\end{figure}

Figures~\ref{Fig:4}a and~\ref{Fig:4}b show CR spectra measured at $T=4.2$~K with both backward wave oscillator (BWO) at 845~GHz~\cite{Q32,Q32a} and quantum cascade laser (QCL) at 3~THz~\cite{Q31} (pulse duration of 3~$\mu$s; repetition period of 100--200 $\mu$s) at different electron concentrations varied by using PPC effect~\cite{SM}. The measurements were performed on the unprocessed sample with a changing of the electron concentration by varying the time illumination from red and blue light emitting diodes placed close to the sample. The GHz/THz radiation passing through the sample was detected either by a silicon bolometer (for BWO) or Ge:Ga photoresistor (for QCL). The electron concentration was determined along with the CR measurements via magnetotransport measurements in the van der Pauw or two-terminal geometry.

All the spectra contain a single CR line, defining the cyclotron mass $m_c$ at the Fermi level. As it is seen from Fig.~\ref{Fig:4}a and~\ref{Fig:4}b, CR lines shift toward high magnetic field with increasing of electron concentration. This indicates that cyclotron mass is strictly increasing function of $n_S$, This fact excludes the gapless state with parabolic band touching in our sample. As it is discussed above, this mass dependence corresponds to the gapless state with the linear band crossing. In order to extract parameters of massless DFs, we have fitted our experimental data by Eq.~(\ref{eq:3}). A good agreement with experimental values is achieved for $A=118$~meV$\cdot$nm, $B=-532$~meV$\cdot$nm$^2$ and $D=-388$~meV$\cdot$nm$^2$. Additionally, we plot the dependencies for $m_c=\hbar^2k_F/A$ (dotted curve) and $m_c=\hbar^2/(2|B|-2D)$ (dashed curve) for pure linear and quadratic band dispersion, respectively.

Figure~\ref{Fig:4}c shows that although there are indeed massless DFs with velocity $v_F=A/\hbar=1.8\cdot10^5$~m/s, they only exist in the immediate vicinity of the $\Gamma$ point at $n_S\leq10^{10}$~cm$^{-2}$, while the terms proportional to $B$ and $D$ are relevant at higher concentrations. Note that additional measurements of temperature dependence of the conductivity at the charge neutrality point also evidences the massless DFs (see the Supplemental Materials~\cite{SM}). Existence of pure massless DFs at small electron concentration is consistent with the absence of odd-sequence of quantum Hall plateaux in magnetotransport of our sample. For instance, for an odd Hall plateau with $5e^2/h$ corresponding to LLs filling factor $\nu=5$ in linear dispersion regime, one should have well-resolved peaks of $R_{xx}$ at magnetic field of 0.08~T. The latter cannot be achieved at the electron mobility of our sample.

Figure~\ref{Fig:4}d compares band structure of the sample numerically calculated within the eight-band Kane model with the analytical expression for $E_{k}(k)$ with the parameters extracted from the fitting $m_c(k_F)$. There is indeed a good agreement for the conduction band, while the valence band is well described at small quasimomentum only. This is also typical for HgTe/CdTe QWs~\cite{Q9,Q10}, in which the discrepancy for the valence band is explained by the effect of the remote subbands beyond the simplified Dirac-like Hamiltonian $H_{\mathrm{D}}(\mathbf{k})$~\cite{Q33}.

In conclusion, we have clearly observed massless DFs in III-V semiconductor QW. Magnetotransport experiments on a gated Hall bar allow us to demonstrate the absence of a band gap in our structure. The  measurements of CR at different concentrations allow us not only determing the velocity $v_F=1.8\cdot10^5$~m/s of the massless DFs, but also demonstrating significant effect of non-linear dispersion at high energies. Experimental dispersion of the massless DFs is in good agreement with realistic band structure calculations based on the eight-band Kane Hamiltonian.


\begin{acknowledgments}
This work was supported by MIPS department of Montpellier University through the "Occitanie Terahertz Platform", by the Languedoc-Roussillon region via the "Gepeto Terahertz platform" and the ARPE project "Terasens" and by the CNRS through "Emergence project 2016" and LIA "TeraMIR". MBE growth of the samples were performed within the French program "Investments for the Future" (ANR-11-EQPX-0016). CR measurements were performed in the framework of Project 17-72-10158 provided by the Russian Science Foundation. S.~S.~Krishtopenko also acknowledges the Ministry of Education and Science of the Russian Federation (MK-1136.2017.2).
\end{acknowledgments}


%

\newpage
\clearpage
\setcounter{equation}{0}
\setcounter{figure}{0}
\setcounter{table}{0}
\renewcommand{\thefigure}{S\arabic{figure}}

\onecolumngrid
\section*{Supplemental Materials}
\maketitle
\onecolumngrid

\subsection{A. Simplified Dirac-like 2D Hamiltonian}
To describe qualitatively electronic states in QWs with inversion symmetry, one can use a simplified Dirac-like Hamiltonian~\cite{sm1}, proposed for the electronic states in \emph{E}1 and \emph{H}1 subbands in the vicinity of the $\Gamma$ point of the Brillouin zone. We note that the given subband is called as an electron-like \emph{E}1 subband, if its electronic states at the $\Gamma$ point of the Brillouin zone are formed by a linear combination of the $|\Gamma_6,\pm1/2\rangle$, $|\Gamma_7,\pm1/2\rangle$ and $|\Gamma_8,\pm1/2\rangle$ bulk bands, while the states in the hole-like \emph{H}1 subband at $\mathbf{k}=0$ have only contribution from the heavy-hole band $|\Gamma_8,\pm3/2\rangle$~\cite{sm7,sm8}. Using the basis states $|E1,+\rangle$, $|H1,+\rangle$, $|E1,-\rangle$, $|H1,-\rangle$, the Hamiltonian for the \emph{E}1 and \emph{H}1 subbands is written as
\begin{equation}\label{eq:SM1}
  H(\mathbf{k}) = \begin{pmatrix}
    H_{D}(\textbf{k}) & 0 \\
    0 & H_{D}^{*}(-\textbf{k})
  \end{pmatrix},
\end{equation}
where
\begin{equation}\label{eq:SM2}
H_{D}(\textbf{k})=\epsilon(\textbf{k})+\sum_{i=1}^3 d_i(\textbf{k})\sigma_i
\end{equation}
and
\begin{equation*}
d_1+id_2=-A(k_x+ik_y)=-Ak_{+},~~~~d_3=M-B(k_x^2+k_y^2),~~~~\epsilon=C-D(k_x^2+k_y^2).
\end{equation*}
Here, $\mathbf{k}=(k_x,k_y)$ are the momentum components in the QW plane, and $A$, $B$, $C$ and $D$ are specific QW constants, being defined by the QW geometry and material parameters. The two components of the Pauli matrices $\sigma_i$ denote the \emph{E}1 and \emph{H}1 subbands, whereas the two diagonal blocks $H_D(\textbf{k})$ and $H_{D}^{*}(-\textbf{k})$ represent spin-up and spin-down states, which are linked together by time-reversal symmetry. Here, as in the main text, we have neglected the terms, arising due to the bulk inversion asymmetry (BIA) of the unit cell~\cite{sm2} and the interface inversion asymmetry (IIA)~\cite{sm3}. The most important quantity in $H(\mathbf{k})$ is the mass parameter $M$, which describes the ordering of \emph{E}1 and \emph{H}1 subbands. At the critical temperature $T=T_c$, the mass parameter is equal to zero. If we then only keep the linear terms in $\textbf{k}$ for each spin, $\hat{H}_{D}^{*}(-\textbf{k})$ and $\hat{H}_{D}(\textbf{k})$ correspond to Hamiltonians, describing massless Dirac fermions. As it has no valley degeneracy, the QW with $M=0$ offers realization of single-valley massless Dirac fermions with the velocity $v_F$ defined by the parameter $A$.

To calculate Landau levels (LLs) in the presence of an external magnetic field $\mathbb{B}$ oriented perpendicular to the QW plane, one should make the Peierls substitution, $\hbar\mathbf{k}\longrightarrow\hbar\mathbf{k}-\frac{e}{c}\mathbf{A}$, where in the Landau gauge $\mathbf{A}=\mathbb{B}(y,0)$. Additionally, we add the Zeeman term in the Hamiltonian
\begin{equation}
\label{eq:SM3}
H_{Z}=\dfrac{1}{2}\mu_B\mathbb{B}\begin{pmatrix}
g_e & 0 & 0 & 0 \\
0 & g_h & 0 & 0 \\
0 & 0 & -g_e & 0 \\
0 & 0 & 0 & -g_h\end{pmatrix},
\end{equation}
where $\mu_B$ is the Bohr magneton, $g_e$ and $g_h$ are the effective (out-of-plane) g-factors of the \emph{E}1 and \emph{H}1 subbands, respectively.

Solving the eigenvalue problem for the upper block of $H(\mathbf{k})+H_{Z}$, the LL energies $E^{(+)}_n$ are found analytically~\cite{sm4}:
\begin{equation*}
E^{(+)}_n=C-\dfrac{2Dn+B}{a_B^2}+\dfrac{g_e+g_h}{4}\mu_B\mathbb{B}
\pm\sqrt{\dfrac{2nA^2}{a_B^2}+\left(M-\dfrac{2Bn+D}{a_B^2}+\dfrac{g_e-g_h}{4}\mu_B\mathbb{B}\right)^2}, \text{~~~~~~~for $n\geq1$}
\end{equation*}
\begin{equation}
\label{eq:SM4}
E^{(+)}_0=C+M-\dfrac{D+B}{a_B^2}+\dfrac{g_e}{2}\mu_B\mathbb{B}, \text{~~~~~~~for $n=0$}.
\end{equation}
For the lower block of $H(\mathbf{k})+H_{Z}$, the LL energies $E^{(-)}_n$ are written as
\begin{equation*}
E^{(-)}_n=C-\dfrac{2Dn-B}{a_B^2}-\dfrac{g_e+g_h}{4}\mu_B\mathbb{B}
\pm\sqrt{\dfrac{2nA^2}{a_B^2}+\left(M-\dfrac{2Bn-D}{a_B^2}-\dfrac{g_e-g_h}{4}\mu_B\mathbb{B}\right)^2}, \text{~~~~~~~for $n\geq1$}
\end{equation*}
\begin{equation}
\label{eq:SM5}
E^{(-)}_0=C-M-\dfrac{D-B}{a_B^2}-\dfrac{g_h}{2}\mu_B\mathbb{B}, \text{~~~~~~~for $n=0$}.
\end{equation}
Here $a_B$ is the magnetic length given by $a_B^2=\hbar c/e\mathbb{B}$. The LLs with energies $E^{(+)}_0$ and $E^{(-)}_0$ are called the \emph{zero-mode} LLs~\cite{sm4}. They split from the edge of \emph{E}1 and \emph{H}1 subbands and tend toward conduction and valence band as a function of $\mathbb{B}$, respectively.

If we put $M$ to zero and omit quadratic terms in the Hamiltonian $H(\mathbf{k})$, Eqs.~(\ref{eq:SM4},\ref{eq:SM5}) are rewritten as
\begin{equation*}
E^{(+)}_n=C+\dfrac{g_e+g_h}{4}\mu_B\mathbb{B}\pm\sqrt{\dfrac{2nA^2}{a_B^2}+\left(\dfrac{g_e-g_h}{4}\mu_B\mathbb{B}\right)^2}, \text{~~~~~~~for $n\geq1$}
\end{equation*}
\begin{equation*}
E^{(+)}_0=C+\dfrac{g_e}{2}\mu_B\mathbb{B}, \text{~~~~~~~for $n=0$}.
\end{equation*}
\begin{equation*}
E^{(-)}_n=C-\dfrac{g_e+g_h}{4}\mu_B\mathbb{B}
\pm\sqrt{\dfrac{2nA^2}{a_B^2}+\left(\dfrac{g_e-g_h}{4}\mu_B\mathbb{B}\right)^2}, \text{~~~~~~~for $n\geq1$}
\end{equation*}
\begin{equation}
\label{eq:SM6}
E^{(-)}_0=C-\dfrac{g_h}{2}\mu_B\mathbb{B}, \text{~~~~~~~for $n=0$}.
\end{equation}
Omitting the Zeeman terms proportional to $g_e$ and $g_h$, we arrive at the spin-degenerate spectrum for LLs in graphene~\cite{sm5} with a square-root dependence of their energies on $\mathbb{B}$.

\begin{figure}
\includegraphics [width=0.72\columnwidth, keepaspectratio] {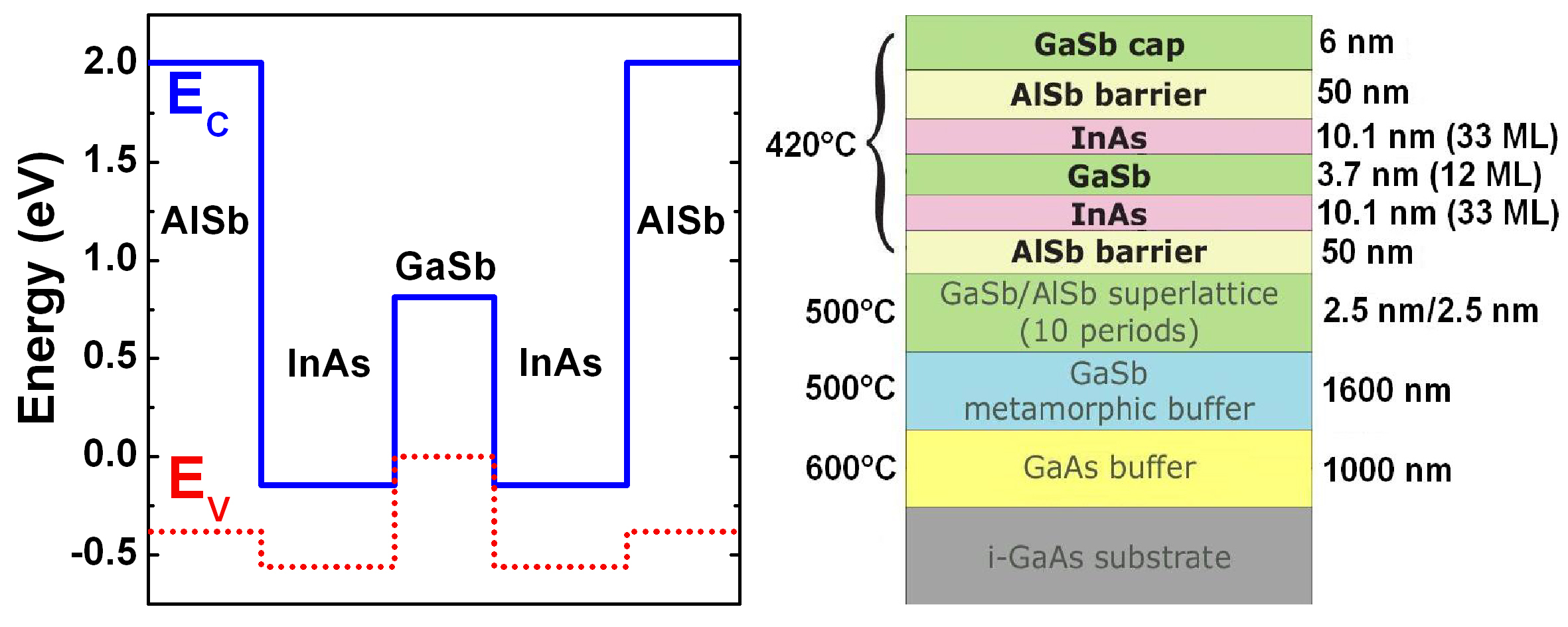} 
\caption{\label{Fig:SM1} Band-edge diagram and growth scheme for three-layer InAs/GaSb/InAs QWs grown on GaSb buffer. The zero energy is referenced to the valence-band edge of bulk GaSb. Outer AlSb barriers provide an overall confining potential for the electron-like and hole-like states.}
\end{figure}

\subsection{B. The sample fabrication}
Figure~\ref{Fig:SM1} schematically shows the band-edge diagram and the growth scheme for three-layer InAs/GaSb/InAs QWs sandwiched between AlSb barriers. The QW studied in this work was grown by solid-source molecular beam epitaxy on a semi-insulating (001) GaAs substrate. After deoxidation at around 600~$^\circ$C a thick undoped GaAs buffer layer was grown. The large lattice-mismatch between GaAs, GaSb and AlSb ($\sim$8\%) was accommodated through a thick GaSb buffer layer of 1.6~$\mu$m followed by a ten-period 2.5 nm GaSb/2.5 nm AlSb superlattice (SLS), all grown at 500~$^\circ$C. Subsequently, the substrate temperature was decreased down to 420~$^\circ$C to grow the three-layer InAs/GaSb/InAs QW confined by 50-nm thick AlSb barrier layers. A 6-nm GaSb cap layer was used to prevent oxidation of the AlSb barrier layers (see Fig.~\ref{Fig:SM1}b). The shutter sequences at all InAs/GaSb interfaces were organized in order to promote the formation of InSb-like interfaces, which gives higher electron mobilities (in contrast to AlAs-like interfaces)~\cite{sm6a}. The barriers on both sides of the QW were nominally undoped. To have a gapless state in the QW, the thicknesses of the InAs and GaSb layers were adjusted to 33~ML and 14~ML, respectively (see Fig.~1 in the main text), where 1~ML corresponds to half of the lattice constant of the bulk material.

The gated Hall bars were fabricated by using a single-mesa process. The fabrication process was defined by standard electron-beam lithography allowing a large scale of size defined between 400~nm and 50~$\mu$m. The process started by the ohmic contact lithography level definition followed by a Pd-based metallization performed by e-beam evaporation. The ohmic contact metallization is then annealed at 300$^\circ$C using a rapid thermal annealing (RTA). Then, 200~nm of SiO$_2$ was deposited by plasma enhanced chemical vapour deposition at 200$^\circ$C. This layer is used as dielectric and allows the fabrication of a gated MOS structure for controlling the carrier density in the QW. The gate lithography was then defined followed by a Ti-based metal gate evaporation. The next step consists in the etching of the SiO$_2$ layer using the gate as a mask. The SiO$_2$ layer is only removed around the Hall bar, the remaining SiO$_2$ being used to isolate the pads from the conductive substrate. To complete the Hall bar, the mesa isolation is done by dry etching using the ohmic contact, gate metal and the SiO$_2$ as a hard mask. Finally, the ohmic contacts and the gate are connected to the pads using air bridges.

\subsection{C. Details of magnetotransport measurements}

Magnetotransport measurements have been performed on 50~$\mu$m wide Hall bars with a low frequency 100 nA ac current. Samples are placed in a variable temperature insert equipped with a $13$~T superconducting coil. Cooling the Hall bars from room temperature down to $T=2$~K reveals a weak insulating behavior as the resistivity doubles. At low temperature the quantum lifetime of the carriers is evaluated from the Shubnikov-de Haas oscillations at low magnetic field and is found equal to $\tau_q=0.17$~ps. The transport scattering time obtained from the mobility is $\tau_{tr}=0.37$~ps, i.e. only 2 times larger than $\tau_q$. An estimate of the LL broadening gives $\Gamma=\hbar/2\tau_{tr}\approx1.8$~meV at zero gate voltage.

\begin{figure}
\includegraphics [width=0.65\columnwidth, keepaspectratio] {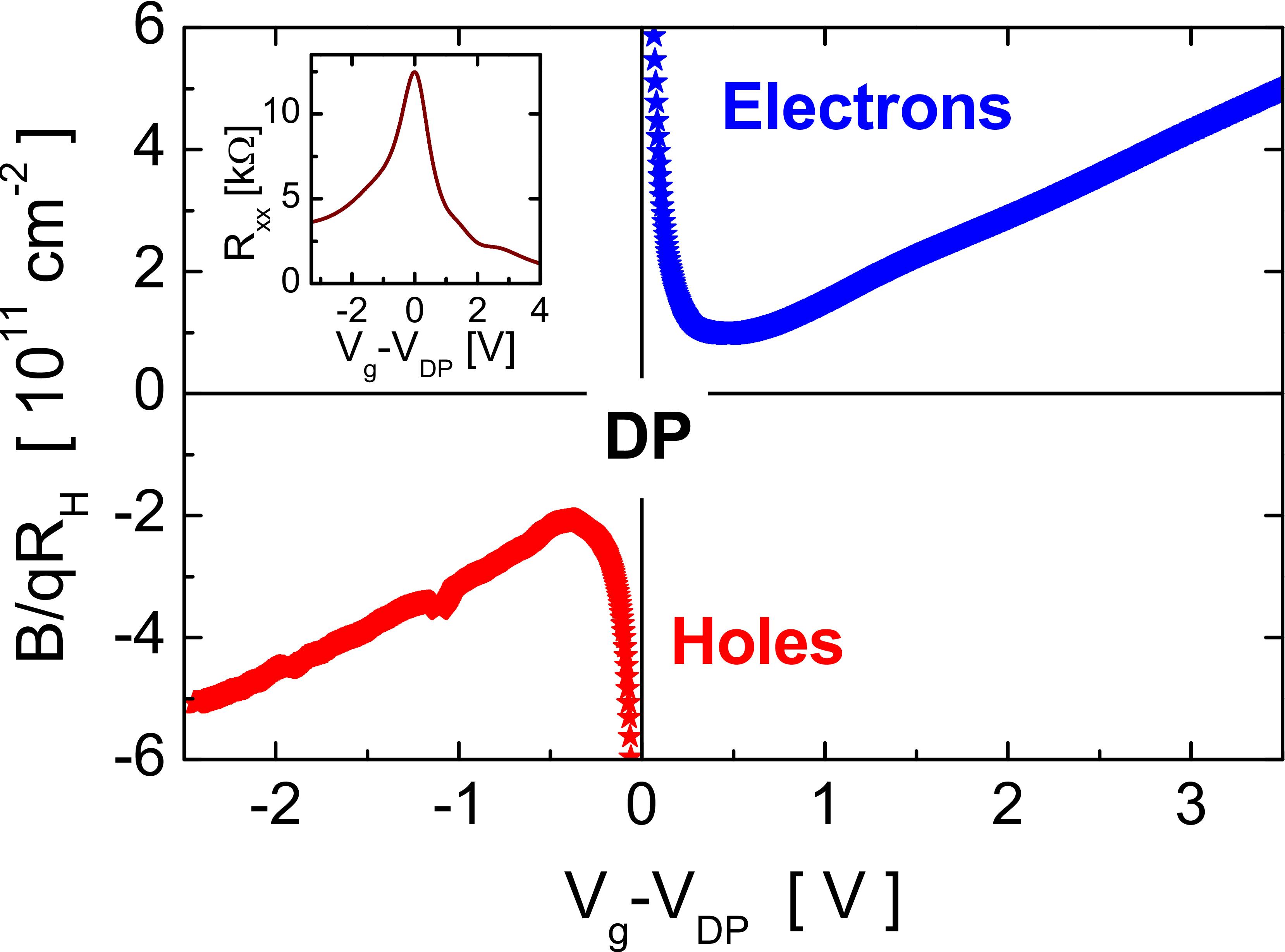} 
\caption{\label{Fig:SM2} Electron and hole densities as a function of gate voltage at $T=1.7$~K and $B=0.2$~T. The inset shows the longitudinal resistance as a function of gate voltage $V_g$ at $B=0$~T.}
\end{figure}

Figure~\ref{Fig:SM2} plots $B/q\mid R_{xy}\mid$, i.e. the electron and hole concentration in a one-carrier approximation at $B=0.2$~T, as a function of gate voltage $V_g$. The carrier density decreases linearly with changing of $V_g-V_{DP}$ from positive to negative values. At the Dirac point $V_{DP}\simeq-4$~V, the data shows an unequivocal inversion of the carrier type versus gate voltage in the QW.

The continuous transition from electrons to holes evidences a gapless state of the QW. At the same time, the longitudinal resistance becomes maximum, when the Fermi level crosses the Dirac point (see the inset in Fig.~\ref{Fig:SM2}). The carrier density variation with gate voltage is similar for electrons and holes with $dn_e /dV_g=1.39\times10^{11}$ ~cm$^{-2}$/V and $dn_h/dV_g=-1.47\times10^{11}$~cm$^{-2}$/V. It means that the conduction and valence band dispersions are almost symmetric in the vicinity of the $\Gamma$ point of the Brillouin zone.

Next, we assume that the carrier density dependence on the gate voltage is controlled mainly by the geometric capacitance of the structure, i.e. $dn/dV_g=C_{geo}/q$. By considering a parallel plate capacitor composed of the silicon oxide layer ($200$~nm), the GaSb cap layer ($6$~nm) and the AlSb barrier ($50$~nm), the equivalent capacitance equals $C_{geo}/A=1.59\times10^{-8}$~F/cm$^2$. It leads to a filling rate of $dn/dV_g= 9.9\times10^{10}$ ~cm$^{-2}$/V which is in agreement with the experimental one. The discrepancy can be explained by the presence of additional charges in the oxide. Indeed a slow drift of the carrier density occurs as the gate is maintained at a fixed negative bias. This effect is minimized during the measurement by scanning the gate voltage back and forth.

\begin{figure}
\includegraphics [width=0.7\columnwidth, keepaspectratio] {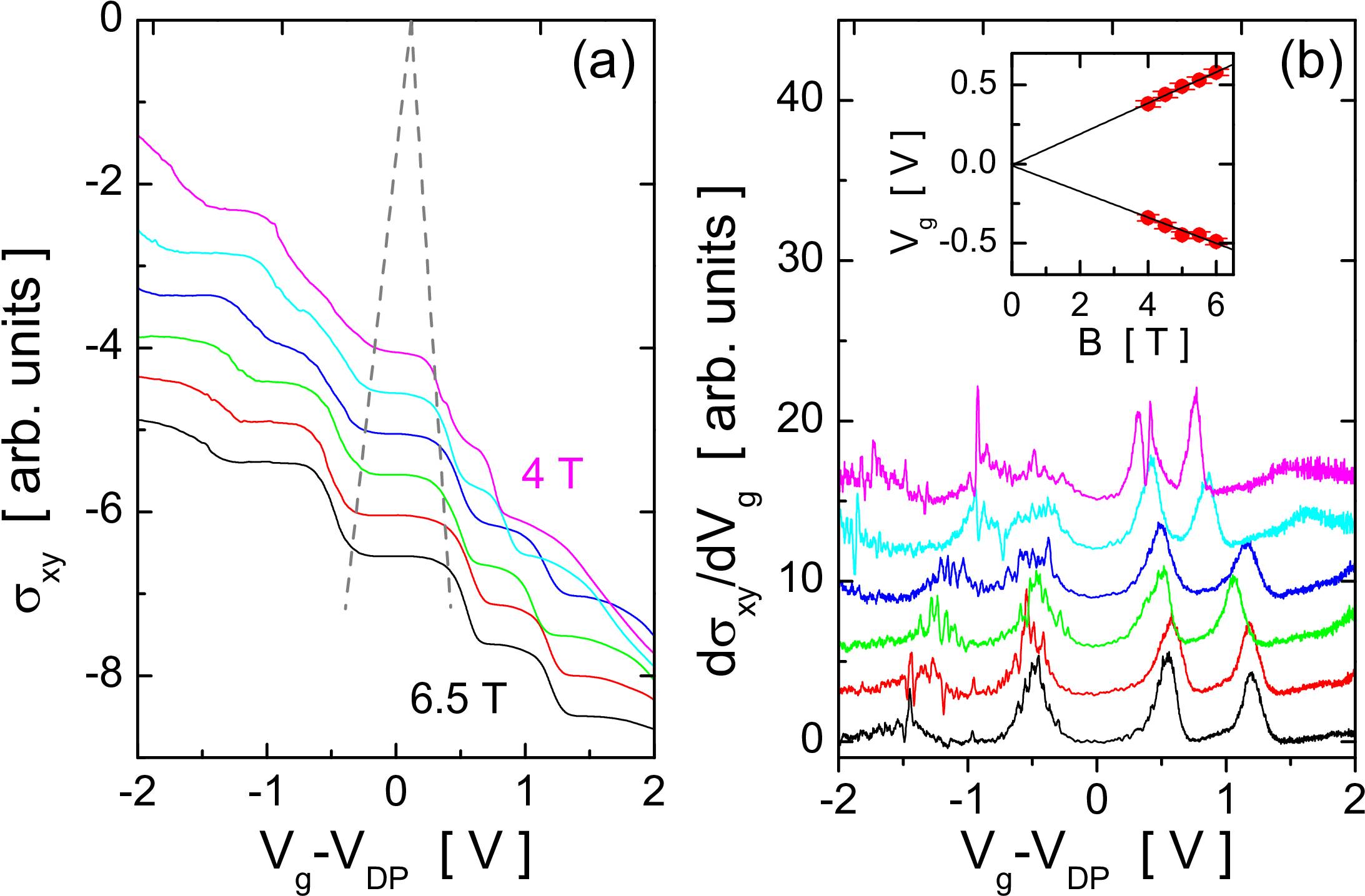} 
\caption{\label{Fig:SM2b} (a) Transverse conductivity vs gate voltage plotted from $B=4$~T to $6.5$~T with a step of $0.5$~T. Curves have been offset for clarity. Dashed lines highlight the shrinkage of the conductivity plateau at the Dirac point when $B$ tends to zero. (b) Derivative $d\sigma_{xy}/dV_g$ plotted vs $V_g$ with a constant offset. Maxima surrounding the Dirac point are clearly visible. They are plotted in the inset as a function of $B$. The linear extrapolation suggests that the gap cancels out at $B=0$~T.}
\end{figure}

Figure~\ref{Fig:SM2b}a shows the transverse conductivity $\sigma_{xy}$ as a function of gate voltage for different magnetic fields. The width of the plateau at zero bias shrinks clearly when $B$ diminishes and tends to zero when $B=0$~T. We underline that only conductivity traces obtained at sufficiently high magnetic fields are plotted for clarity. The $\sigma_{xy}$ plateau is strongly smoothed at lower fields. A reliable method for extracting the critical field correctly is to plot the maxima in the $d\sigma_{xy}/dV_g(V_g)$ curves as a function of $B$ (see Fig.~\ref{Fig:SM2b}b and its inset). The linear extrapolation indicates that the gap cancels out at $B\sim0$~T, which confirms the critical field value measured from the $R_{xx}=e^2/h$ criterion reported in the main text.

\subsection{D. Temperature dependence of conductivity at charge neutrality point}

\begin{figure}
\includegraphics [width=0.85\columnwidth, keepaspectratio] {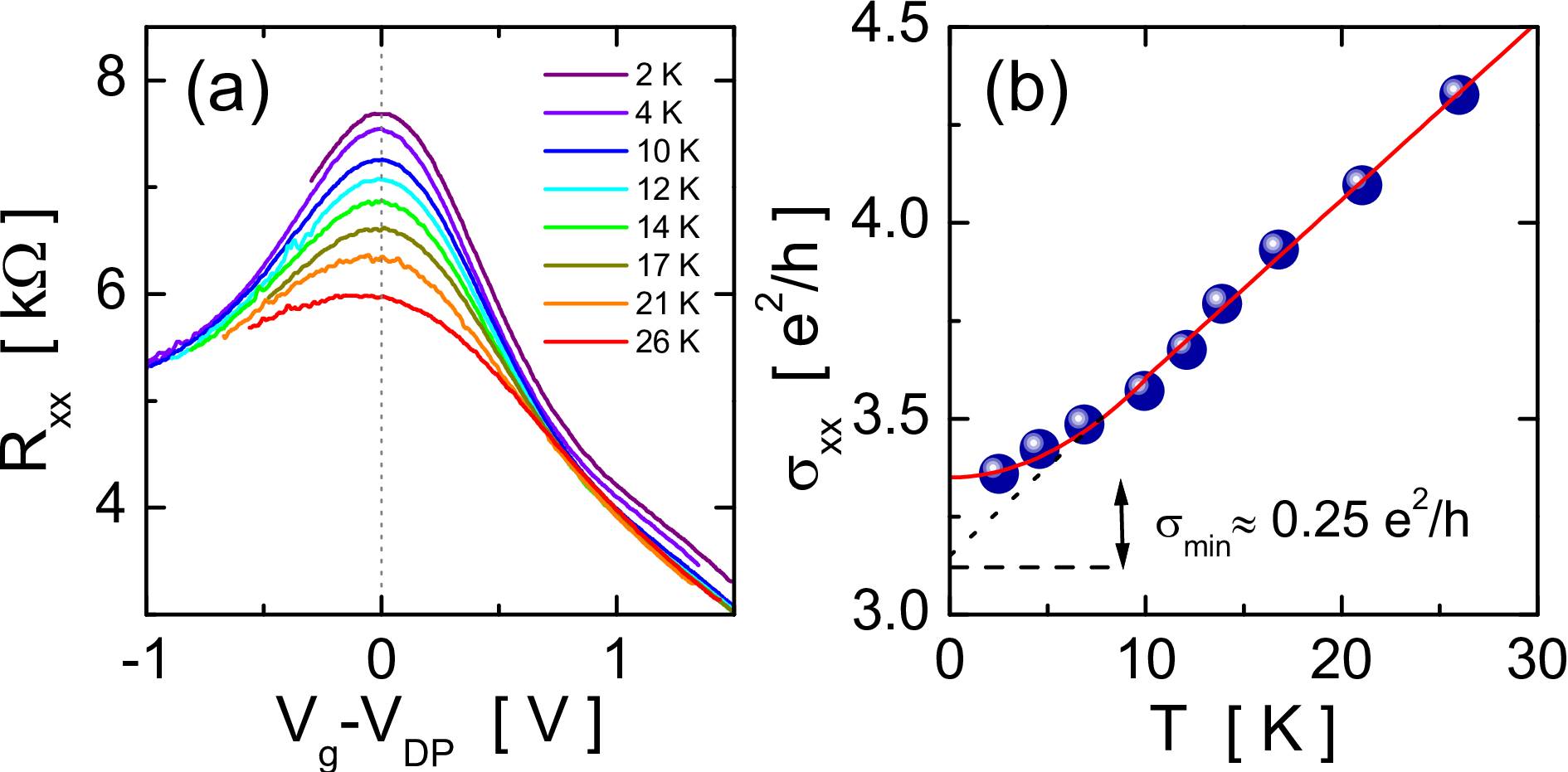} 
\caption{\label{Fig:SM3} (a) Zero-field resistance $R_{xx}$ as a function of swing voltage $V_g-V_{DP}$ for different temperatures. (b) Conductivity at the Dirac point as a function of temperature. Solid circles are experimental data. The red line is a fit to exact expression of the conductivity within the Dirac-like model based on the Kubo formula derived in~\cite{smQ6a}. The black dashed line represents the high temperature behavior described by Eq.~(\ref{eq:SM7b}).}
\end{figure}

Figure~\ref{Fig:SM3}a plots the resistance $R_{xx}$ versus swing voltage $V_g-V_{DP}$ at different temperatures measured on a second Hall bar. The graph clearly shows the peaked resistivity behavior expected for Dirac-like 2D systems~\cite{smQ6a,sm10,sm11}. As can be seen, the maximum of $R_{xx}$ decreases with $T$. Figure~\ref{Fig:SM3}b shows the conductivity at the Dirac point as a function of temperature.
Qualitatively, the temperature dependence of $\sigma_{xx}$ is written as~\cite{smQ6a}:
\begin{equation}
\label{eq:SM7a}
\sigma_{xx}\approx \dfrac{2}{\pi}\dfrac{e^2}{h}\dfrac{1}{1+\langle M^2\rangle/\Gamma^2}+\mathcal{O}\left(T^2\right), \text{~~~~$k_BT\ll\Gamma$,}
\end{equation}
\begin{equation}
\label{eq:SM7b}
\sigma_{xx}\propto \dfrac{e^2}{h}\dfrac{k_BT}{\Gamma}, \text{~~~~$k_BT\geq\Gamma$,}
\end{equation}
where $\Gamma$ is the spectral broadening induced by spin-independent potential disorder and $\langle M^2\rangle$ is the variance of the gap due to spatial deviations of the layer thicknesses from their critical values.

As seen from Fig.~\ref{Fig:SM3}b, the conductivity shows a quadratic dependence on $T$ at low temperature with the minimal value $\sigma_{xx}>3\,e^2/h$. This large value is caused by partial conduction via the sample bulk, which is typical for the InAs/GaSb QWs~\cite{sm9}. In order to estimate the residual conductivity, the linear dependence of $\sigma_{xx}$ at high temperature is extrapolated down to $T=0$~K. By subtracting the obtained value to $\sigma_{xx}(0)$ we find $\sigma_{\text{min}}\approx0.25\,e^2/h$. As clear from Eq.~(\ref{eq:SM7b}), the conductivity slope at high $T$ is inversely proportional to the broadening parameter $\Gamma$ induced by the disorder. By fitting the high-temperature values, we obtain $\Gamma\approx 1.9$~meV, which is in a good agreement with the LL broadening of $\Gamma=\hbar/2\tau_{tr}\approx1.8$~meV extracted from magnetotransport (see Fig.~2 in the main text and Sec.~C). From the values of $\sigma_\text{min}$ and $\Gamma$, we also evaluate the gap fluctuation in our sample $\sqrt{\langle\mathcal{M}^2\rangle}\approx 1.5 \,\Gamma \approx 3$~meV. We note that the linear temperature dependence of $\sigma_{xx}$ at high $T$ resulting from the linear density of states is another manifestation of the DFs in the QW~\cite{smQ6a}.

\subsection{E. Quadratic band touching in zinc-blende QWs}
Let us now consider various cases of quadratic band touching at the $\Gamma$ point of Brillouin zone, which may exist in symmetric zinc-blende QWs grown in the [001] direction. The form of effective Hamiltonian for quadratic band touching can be obtained from the symmetry considerations. For simplicity, we neglect the terms, arising due to the bulk inversion asymmetry (BIA) of the unit cell~\cite{sm2} and the interface inversion asymmetry (IIA)~\cite{sm3}.

In this case, for electronic QW states at the $\Gamma$ point with in-plane momentum $k=0$, we have three types of symmetries: the time reversal symmetry, the inversion symmetry and the in-plane full rotation symmetry~\cite{sm12,smQ16,sm13}. Time reversal symmetry relates states with opposite spin to each other; hence when the effective Hamiltonian for one spin is constructed, the Hamiltonian for the opposite spin can be easily obtained through the time-reversal operation. The inversion operation defines the parity of each subband, while in-plane rotation symmetry allows for conservation of the total angular momentum $J$ along the growth QW direction. Both inversion and  in-plane full rotation symmetries define the form of matrix elements in the effective Hamiltonian. From the symmetry considerations mentioned above, the general form of the effective Hamiltonian describing quadratic band touching is
\begin{equation}\label{eq:SM8}
  \mathcal{H}_{k^2}(\mathbf{k}) = \begin{pmatrix}
    H_{k^2}(\mathbf{k}) & 0 \\
    0 & H_{k^2}^{*}(-\textbf{k})
  \end{pmatrix}
\end{equation}
where $H_{k^2}(\mathbf{k})$ has the form similar to $H_{D}(\textbf{k})$ in Eq.~(\ref{eq:SM2}):
\begin{equation}\label{eq:SM9}
  H_{k^2}(\mathbf{k}) = \epsilon(\textbf{k})+\begin{pmatrix}
    d_3(\textbf{k}) & d_1(\textbf{k})+id_2(\textbf{k}) \\
    d_1(\textbf{k})-id_2(\textbf{k}) & -d_3(\textbf{k})
  \end{pmatrix}.
\end{equation}
As we intend to keep in $\mathcal{H}_{k^2}(\mathbf{k})$ only the terms up to \emph{the third order} in $k$, $\epsilon(\textbf{k})$ and $d_3(\textbf{k})$ can be written as
\begin{equation}
\label{eq:SM10}
\epsilon=C-Dk^2,~~~~~~~~~~~d_3=-Bk^2,
\end{equation}
where $k^2=k_x^2+k_y^2.$

First, we consider the case of different parity of the conduction and valence band. Away from the $\Gamma$ point, these states can mix. However, the coupling matrix element between these two states must be an odd function of the in-plane momentum $k$ and obey full rotation symmetry. Thus, we deduce two forms of $d_1(\textbf{k})+id_2(\textbf{k})$ in $H_{k^2}(\mathbf{k})$:
\begin{equation}
\label{eq:SM11}
d_1+id_2=-\mathcal{A}k^2k_{+},~~~~~~~~~~~d_1+id_2=-\mathcal{A}k_{+}^3.
\end{equation}
The former can be realized in 2D system, described by $H_{D}(\textbf{k})$ in the absence of linear term $Ak_{\pm}$~\cite{sm13}, while the latter takes place in double HgTe QWs~\cite{sm12} or GaSb/InAs/GaSb QWs~\cite{smQ16}. The energy dispersion for these two cases has the same form:
\begin{equation}
\label{eq:SM12}
E_{k^2}(k)=C-Dk^2\pm k^2\sqrt{B^2+\mathcal{A}^2k^2},
\end{equation}
where "$+$" and "$-$" correspond to the conduction and valence bands, respectively.

In the opposite case of the same parity of conduction and valence band, the coupling matrix element must be an even function of $k$. Taking into account the full rotation symmetry, the non-zero $d_1(\textbf{k})+id_2(\textbf{k})$ is written as:
\begin{equation}
\label{eq:SM13}
d_1+id_2=-\mathcal{B}k_{+}^2,
\end{equation}
resulting to the following band dispersion:
\begin{equation}
\label{eq:SM14}
E_{k^2}(k)=C-Dk^2\pm k^2\sqrt{B^2+\mathcal{B}^2}.
\end{equation}
As it is seen that in terms of band dispersion, the interband mixing formally results in renormalization of $B$, which is now written as $\sqrt{B^2+\mathcal{B}^2}$.

Formally, we can rewrite band dispersions in Eq.~(\ref{eq:SM12}) and (\ref{eq:SM14}) in a joint form
\begin{equation*}
E_{k^2}(k)=C-Dk^2\pm k^2\sqrt{B^2+\mathcal{A}^2k^2},
\end{equation*}
in which $\mathcal{A}\neq0$ corresponds to the opposite parity of conduction and valence band, while for the same parity, one should set $\mathcal{A}=0$.

\subsection{F. Persistent photoconductivity effect}

Persistent photoconductivity (PPC) is the phenomenon of long-term modification of the material conductivity after the action of light at low temperatures. If the sample conductivity increases after illumination, the effect is called \emph{positive} PPC. In the case of conductivity decreasing after the action of light, PPC is named as \emph{negative}. The inherent property of the InAs-based QWs is the bipolarity of PPC, arising under illumination of the sample at different wavelengths~\cite{sm6,sm23,sm25,sm26,sm28,sm28a}. The negative PPC was first observed in~\cite{sm6} under the illumination by green light emitting diode (LED), while the positive PPC was measured in \cite{sm23} under the influence of infrared (IR) light. By using bipolar PPC, the electron concentration can be reversibly changed by several times, which offers tuning the Fermi level in the sample without a gate.

As we are interested in the states close in energy to the band crossing point (see the main text), we first determine the wavelength of the LED, providing the most efficient way to decrease the electron concentration in our sample. For such purpose, we perform a spectral study of PPC effect. The scheme of experimental setup is based on MDR-23 grating monochromator~\cite{sm25}.
A quartz incandescent lamp was used as the radiation source, and the higher-order diffraction peaks were cut off by standard filters. At the monochromator output, radiation with photon energy from 0.6 to 4 eV was coupled to an optical fiber and delivered to the sample holder in helium cryostat. In the PPC study, we used the rectangular samples of the Hall geometry with two strip indium ohmic contacts deposited at the edges. A dc current of 1~$\mu$A was passed through the sample placed at the center of a superconducting solenoid. All the measurements were performed at $T=4.2$~K.

The photoconductivity spectra were recorded in two different modes. In the first mode the measurements were performed step by step starting from the long-wavelength part of the spectrum. The sample was illuminated with monochromatic radiation then after switching off the illumination, to restore the "equilibrium" value of the resistance, the exposure was performed (typically a few tens of seconds) and magnetotransport measurements were carried out. Then the monochromator was re-tuned to a shorter wavelength and the procedure repeated. To determine the electron concentration, we also measured resistances as a function of magnetic field in a two-stripe geometry. In the second mode, the measurements were performed under continuous illumination of monochromatic radiation with the wavelength slowly scanned starting from the short-wavelength part of the spectrum. The scanning pitch on the wavelength was 2~nm in the short-wavelength part and~20 nm in the long-wavelength part of the spectrum. After each step the signal was averaged during time which in our measurements was typically 20 seconds. The typical time of a spectral recording was several hours.

\begin{figure}
\includegraphics [width=0.95\columnwidth, keepaspectratio] {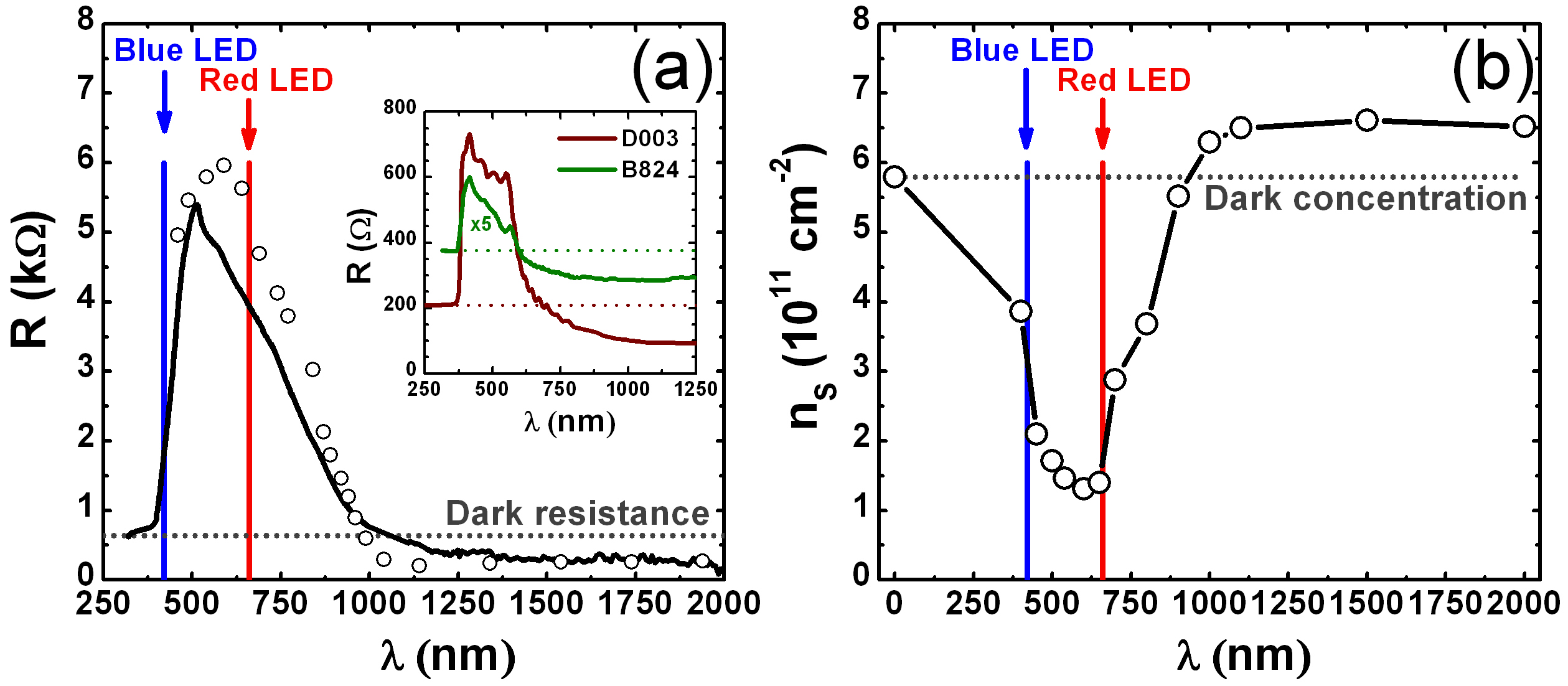} 
\caption{\label{Fig:SM4} (a) PPC spectrum of our sample. The symbol and solid curve correspond to different methods of the measurements. The dotted line shows the dark resistance of the sample. The insets provides PPC spectra for single InAs/AlSb QW of 15 nm width (B824)~\cite{sm25} and a 15 nm wide double InAs/AlSb QW with the middle AlSb barrier of 5~nm thickness (D003)~\cite{sm26}. (b) The maximum electron concentration achieved at the given wavelength. The dotted line corresponds to the dark value measured after cooling and before the first illumination. The vertical lines marked the wavelength of LED used in the measurements of cyclotron resonance.}
\end{figure}

Figure~\ref{Fig:SM4}a represents the PPC spectrum for our sample. For comparison, the inset also provides PPC spectra for the single InAs/AlSb QW of 15 nm width (B824)~\cite{sm25} and a 15 nm wide double InAs/AlSb QW with the middle AlSb barrier of 5~nm thickness (D003)~\cite{sm26}, both without GaSb layer in the QW. In the long-wavelength part of the spectrum at $\lambda>1000$~nm, the positive PPC is observed (the resistance is lower than the dark one). In the range between 300 and 1000~nm, a pronounced negative PPC is clearly seen. The measurements of maximum electron concentration achieved at the given wavelength (when the sample resistance saturates) show a clear correlation between spectral dependencies of the  resistance and electron concentration (see Fig.~\ref{Fig:SM4}b). In the spectral range of positive PPC an increase of electron concentration (compared to the dark value) is seen, while in the range of negative PPC the concentration decreases. Thus, PPC effect can be indeed used as an optical gating for the sample.

By comparing the PPC spectrum of our sample with the spectra of single and double InAs/AlSb QWs shown in the inset, we conclude that the main spectral features of PPC in all the cases coincide. It means that the origin of PPC in our sample and in InAs/AlSb QWs should be the same. The differences in the width of the negative PPC range are caused by the presence of the GaSb layer in the QWs. As it is recently shown for the InAs/Ga(In)Sb QW bilayers~\cite{sm28}, it affects a lot the QW response to the IR illumination.

Both positive and negative PPC are caused by recharge of the deep donors in the GaSb cap layer~\cite{sm23,sm25,sm26}. As it was shown in~\cite{sm23,sm25}, the negative PPC is mainly associated with the charge transfer from the InAs QW to the charged surface donors in the cap GaSb layer. The negative PPC is related with band gap excitation of electron-hole pairs firstly in the GaSb cap layer and then (at shorter wavelengths) in the AlSb barriers. Photoexcited electrons are captured by ionized deep donors, while the holes drift in the AlSb barriers in the "built-in" electric field to the interface AlSb/InAs, where they recombine with electrons in InAs QW, thus, decreasing 2D electron concentration.

Since the positive and negative PPC are reversible (i.e. by sequentially illuminating the sample by visible and IR radiation we can reversibly vary the 2D electron concentration in the structure), then, both positive and negative PPC should be finally associated with recharge of the same deep centers. As shown in~\cite{sm26}, the transfer of the electric charge from the GaSb cap layer under the IR illumination  has a "diffusive" character and is apparently performed via the impurity states in the AlSb barrier. Since the electron effective mass is smaller than the hole effective mass, electrons "diffuse" more rapidly. The holes providing an excess of positive charges in the cap layer under the band-to-band illumination in GaSb will be captured by neutral surface donors in this case~\cite{sm26}. Note that the infrared photon energy is high enough to ionize the deep donors in the AlSb barriers, which also contributes into positive PPC effect.

In the cyclotron resonance measurements, we used both blue and red LEDs. However, illumination by the red LED resulted in a wider range of concentration than with the blue LED.



\begin{thebibliography}{51}%
\makeatletter
\providecommand \@ifxundefined [1]{%
 \@ifx{#1\undefined}
}%
\providecommand \@ifnum [1]{%
 \ifnum #1\expandafter \@firstoftwo
 \else \expandafter \@secondoftwo
 \fi
}%
\providecommand \@ifx [1]{%
 \ifx #1\expandafter \@firstoftwo
 \else \expandafter \@secondoftwo
 \fi
}%
\providecommand \natexlab [1]{#1}%
\providecommand \enquote  [1]{``#1''}%
\providecommand \bibnamefont  [1]{#1}%
\providecommand \bibfnamefont [1]{#1}%
\providecommand \citenamefont [1]{#1}%
\providecommand \href@noop [0]{\@secondoftwo}%
\providecommand \href [0]{\begingroup \@sanitize@url \@href}%
\providecommand \@href[1]{\@@startlink{#1}\@@href}%
\providecommand \@@href[1]{\endgroup#1\@@endlink}%
\providecommand \@sanitize@url [0]{\catcode `\\12\catcode `\$12\catcode
  `\&12\catcode `\#12\catcode `\^12\catcode `\_12\catcode `\%12\relax}%
\providecommand \@@startlink[1]{}%
\providecommand \@@endlink[0]{}%
\providecommand \url  [0]{\begingroup\@sanitize@url \@url }%
\providecommand \@url [1]{\endgroup\@href {#1}{\urlprefix }}%
\providecommand \urlprefix  [0]{URL }%
\providecommand \Eprint [0]{\href }%
\providecommand \doibase [0]{http://dx.doi.org/}%
\providecommand \selectlanguage [0]{\@gobble}%
\providecommand \bibinfo  [0]{\@secondoftwo}%
\providecommand \bibfield  [0]{\@secondoftwo}%
\providecommand \translation [1]{[#1]}%
\providecommand \BibitemOpen [0]{}%
\providecommand \bibitemStop [0]{}%
\providecommand \bibitemNoStop [0]{.\EOS\space}%
\providecommand \EOS [0]{\spacefactor3000\relax}%
\providecommand \BibitemShut  [1]{\csname bibitem#1\endcsname}%
\let\auto@bib@innerbib\@empty
\bibitem [{\citenamefont {Novoselov}\ \emph {et~al.}(2005)\citenamefont
  {Novoselov}, \citenamefont {Geim}, \citenamefont {Morozov}, \citenamefont
  {Jiang}, \citenamefont {Katsnelson}, \citenamefont {Grigorieva},
  \citenamefont {Dubonos},\ and\ \citenamefont {Firsov}}]{Q1}%
  \BibitemOpen
  \bibfield  {author} {\bibinfo {author} {\bibfnamefont {K.~S.}\ \bibnamefont
  {Novoselov}}, \bibinfo {author} {\bibfnamefont {A.~K.}\ \bibnamefont {Geim}},
  \bibinfo {author} {\bibfnamefont {S.~V.}\ \bibnamefont {Morozov}}, \bibinfo
  {author} {\bibfnamefont {D.}~\bibnamefont {Jiang}}, \bibinfo {author}
  {\bibfnamefont {M.~I.}\ \bibnamefont {Katsnelson}}, \bibinfo {author}
  {\bibfnamefont {I.~V.}\ \bibnamefont {Grigorieva}}, \bibinfo {author}
  {\bibfnamefont {S.~V.}\ \bibnamefont {Dubonos}}, \ and\ \bibinfo {author}
  {\bibfnamefont {A.~A.}\ \bibnamefont {Firsov}},\ }\href {\doibase
  10.1038/nature04233} {\bibfield  {journal} {\bibinfo  {journal} {Nature}\
  }\textbf {\bibinfo {volume} {438}},\ \bibinfo {pages} {197} (\bibinfo {year}
  {2005})}\BibitemShut {NoStop}%
\bibitem [{\citenamefont {Wehling}\ \emph {et~al.}(2014)\citenamefont
  {Wehling}, \citenamefont {Black-Schaffer},\ and\ \citenamefont
  {Balatsky}}]{Q2}%
  \BibitemOpen
  \bibfield  {author} {\bibinfo {author} {\bibfnamefont {T.~O.}\ \bibnamefont
  {Wehling}}, \bibinfo {author} {\bibfnamefont {A.~M.}\ \bibnamefont
  {Black-Schaffer}}, \ and\ \bibinfo {author} {\bibfnamefont {A.~V.}\
  \bibnamefont {Balatsky}},\ }\href {\doibase 10.1080/00018732.2014.927109}
  {\bibfield  {journal} {\bibinfo  {journal} {Adv. Phys.}\ }\textbf {\bibinfo
  {volume} {63}},\ \bibinfo {pages} {1} (\bibinfo {year} {2014})}\BibitemShut
  {NoStop}%
\bibitem [{\citenamefont {Castro~Neto}\ \emph {et~al.}(2009)\citenamefont
  {Castro~Neto}, \citenamefont {Guinea}, \citenamefont {Peres}, \citenamefont
  {Novoselov},\ and\ \citenamefont {Geim}}]{Q3}%
  \BibitemOpen
  \bibfield  {author} {\bibinfo {author} {\bibfnamefont {A.~H.}\ \bibnamefont
  {Castro~Neto}}, \bibinfo {author} {\bibfnamefont {F.}~\bibnamefont {Guinea}},
  \bibinfo {author} {\bibfnamefont {N.~M.~R.}\ \bibnamefont {Peres}}, \bibinfo
  {author} {\bibfnamefont {K.~S.}\ \bibnamefont {Novoselov}}, \ and\ \bibinfo
  {author} {\bibfnamefont {A.~K.}\ \bibnamefont {Geim}},\ }\href {\doibase
  10.1103/RevModPhys.81.109} {\bibfield  {journal} {\bibinfo  {journal} {Rev.
  Mod. Phys.}\ }\textbf {\bibinfo {volume} {81}},\ \bibinfo {pages} {109}
  (\bibinfo {year} {2009})}\BibitemShut {NoStop}%
\bibitem [{\citenamefont {Beenakker}(2008)}]{Q1c}%
  \BibitemOpen
  \bibfield  {author} {\bibinfo {author} {\bibfnamefont {C.~W.~J.}\
  \bibnamefont {Beenakker}},\ }\href {\doibase 10.1103/RevModPhys.80.1337}
  {\bibfield  {journal} {\bibinfo  {journal} {Rev. Mod. Phys.}\ }\textbf
  {\bibinfo {volume} {80}},\ \bibinfo {pages} {1337} (\bibinfo {year}
  {2008})}\BibitemShut {NoStop}%
\bibitem [{\citenamefont {Liang}\ \emph {et~al.}(2015)\citenamefont {Liang},
  \citenamefont {Gibson}, \citenamefont {Ali}, \citenamefont {Liu},
  \citenamefont {Cava},\ and\ \citenamefont {Ong}}]{Q1d}%
  \BibitemOpen
  \bibfield  {author} {\bibinfo {author} {\bibfnamefont {T.}~\bibnamefont
  {Liang}}, \bibinfo {author} {\bibfnamefont {Q.}~\bibnamefont {Gibson}},
  \bibinfo {author} {\bibfnamefont {M.~N.}\ \bibnamefont {Ali}}, \bibinfo
  {author} {\bibfnamefont {M.}~\bibnamefont {Liu}}, \bibinfo {author}
  {\bibfnamefont {R.~J.}\ \bibnamefont {Cava}}, \ and\ \bibinfo {author}
  {\bibfnamefont {N.~P.}\ \bibnamefont {Ong}},\ }\href {\doibase
  10.1038/nmat4143} {\bibfield  {journal} {\bibinfo  {journal} {Nat. Mater.}\
  }\textbf {\bibinfo {volume} {14}},\ \bibinfo {pages} {280Ц284} (\bibinfo
  {year} {2015})}\BibitemShut {NoStop}%
\bibitem [{\citenamefont {Novoselov}\ \emph {et~al.}(2004)\citenamefont
  {Novoselov}, \citenamefont {Geim}, \citenamefont {Morozov}, \citenamefont
  {Jiang}, \citenamefont {Zhang}, \citenamefont {Dubonos}, \citenamefont
  {Grigorieva},\ and\ \citenamefont {Firsov}}]{Q1b}%
  \BibitemOpen
  \bibfield  {author} {\bibinfo {author} {\bibfnamefont {K.~S.}\ \bibnamefont
  {Novoselov}}, \bibinfo {author} {\bibfnamefont {A.~K.}\ \bibnamefont {Geim}},
  \bibinfo {author} {\bibfnamefont {S.~V.}\ \bibnamefont {Morozov}}, \bibinfo
  {author} {\bibfnamefont {D.}~\bibnamefont {Jiang}}, \bibinfo {author}
  {\bibfnamefont {Y.}~\bibnamefont {Zhang}}, \bibinfo {author} {\bibfnamefont
  {S.~V.}\ \bibnamefont {Dubonos}}, \bibinfo {author} {\bibfnamefont {I.~V.}\
  \bibnamefont {Grigorieva}}, \ and\ \bibinfo {author} {\bibfnamefont {A.~A.}\
  \bibnamefont {Firsov}},\ }\href {\doibase 10.1126/science.1102896} {\bibfield
   {journal} {\bibinfo  {journal} {Science}\ }\textbf {\bibinfo {volume}
  {306}},\ \bibinfo {pages} {666} (\bibinfo {year} {2004})}\BibitemShut
  {NoStop}%
\bibitem [{\citenamefont {Gerchikov}\ and\ \citenamefont
  {Subashiev}(1990)}]{Q4}%
  \BibitemOpen
  \bibfield  {author} {\bibinfo {author} {\bibfnamefont {L.~G.}\ \bibnamefont
  {Gerchikov}}\ and\ \bibinfo {author} {\bibfnamefont {A.~V.}\ \bibnamefont
  {Subashiev}},\ }\href {\doibase 10.1002/pssb.2221600207} {\bibfield
  {journal} {\bibinfo  {journal} {Phys. Status Solidi B}\ }\textbf {\bibinfo
  {volume} {160}},\ \bibinfo {pages} {443} (\bibinfo {year}
  {1990})}\BibitemShut {NoStop}%
\bibitem [{\citenamefont {Bernevig}\ \emph {et~al.}(2006)\citenamefont
  {Bernevig}, \citenamefont {Hughes},\ and\ \citenamefont {Zhang}}]{Q5}%
  \BibitemOpen
  \bibfield  {author} {\bibinfo {author} {\bibfnamefont {B.~A.}\ \bibnamefont
  {Bernevig}}, \bibinfo {author} {\bibfnamefont {T.~L.}\ \bibnamefont
  {Hughes}}, \ and\ \bibinfo {author} {\bibfnamefont {S.-C.}\ \bibnamefont
  {Zhang}},\ }\href {\doibase 10.1126/science.1133734} {\bibfield  {journal}
  {\bibinfo  {journal} {Science}\ }\textbf {\bibinfo {volume} {314}},\ \bibinfo
  {pages} {1757} (\bibinfo {year} {2006})}\BibitemShut {NoStop}%
\bibitem [{\citenamefont {B\"{u}ttner}\ \emph {et~al.}(2011)\citenamefont
  {B\"{u}ttner}, \citenamefont {Liu}, \citenamefont {Tkachov}, \citenamefont
  {Novik}, \citenamefont {Br\"{u}ne}, \citenamefont {Buhmann}, \citenamefont
  {Hankiewicz}, \citenamefont {Recher}, \citenamefont {Trauzettel},
  \citenamefont {Zhang},\ and\ \citenamefont {Molenkamp}}]{Q6a}%
  \BibitemOpen
  \bibfield  {author} {\bibinfo {author} {\bibfnamefont {B.}~\bibnamefont
  {B\"{u}ttner}}, \bibinfo {author} {\bibfnamefont {C.}~\bibnamefont {Liu}},
  \bibinfo {author} {\bibfnamefont {G.}~\bibnamefont {Tkachov}}, \bibinfo
  {author} {\bibfnamefont {E.}~\bibnamefont {Novik}}, \bibinfo {author}
  {\bibfnamefont {C.}~\bibnamefont {Br\"{u}ne}}, \bibinfo {author}
  {\bibfnamefont {H.}~\bibnamefont {Buhmann}}, \bibinfo {author} {\bibfnamefont
  {E.}~\bibnamefont {Hankiewicz}}, \bibinfo {author} {\bibfnamefont
  {P.}~\bibnamefont {Recher}}, \bibinfo {author} {\bibfnamefont
  {B.}~\bibnamefont {Trauzettel}}, \bibinfo {author} {\bibfnamefont
  {S.}~\bibnamefont {Zhang}}, \ and\ \bibinfo {author} {\bibfnamefont
  {L.}~\bibnamefont {Molenkamp}},\ }\href {\doibase 10.1038/nphys1914}
  {\bibfield  {journal} {\bibinfo  {journal} {Nat. Phys.}\ }\textbf {\bibinfo
  {volume} {7}},\ \bibinfo {pages} {418} (\bibinfo {year} {2011})}\BibitemShut
  {NoStop}%
\bibitem [{\citenamefont {Zholudev}\ \emph {et~al.}(2012)\citenamefont
  {Zholudev}, \citenamefont {Teppe}, \citenamefont {Orlita}, \citenamefont
  {Consejo}, \citenamefont {Torres}, \citenamefont {Dyakonova}, \citenamefont
  {Czapkiewicz}, \citenamefont {Wr\'obel}, \citenamefont {Grabecki},
  \citenamefont {Mikhailov}, \citenamefont {Dvoretskii}, \citenamefont
  {Ikonnikov}, \citenamefont {Spirin}, \citenamefont {Aleshkin}, \citenamefont
  {Gavrilenko},\ and\ \citenamefont {Knap}}]{Q6b}%
  \BibitemOpen
  \bibfield  {author} {\bibinfo {author} {\bibfnamefont {M.}~\bibnamefont
  {Zholudev}}, \bibinfo {author} {\bibfnamefont {F.}~\bibnamefont {Teppe}},
  \bibinfo {author} {\bibfnamefont {M.}~\bibnamefont {Orlita}}, \bibinfo
  {author} {\bibfnamefont {C.}~\bibnamefont {Consejo}}, \bibinfo {author}
  {\bibfnamefont {J.}~\bibnamefont {Torres}}, \bibinfo {author} {\bibfnamefont
  {N.}~\bibnamefont {Dyakonova}}, \bibinfo {author} {\bibfnamefont
  {M.}~\bibnamefont {Czapkiewicz}}, \bibinfo {author} {\bibfnamefont
  {J.}~\bibnamefont {Wr\'obel}}, \bibinfo {author} {\bibfnamefont
  {G.}~\bibnamefont {Grabecki}}, \bibinfo {author} {\bibfnamefont
  {N.}~\bibnamefont {Mikhailov}}, \bibinfo {author} {\bibfnamefont
  {S.}~\bibnamefont {Dvoretskii}}, \bibinfo {author} {\bibfnamefont
  {A.}~\bibnamefont {Ikonnikov}}, \bibinfo {author} {\bibfnamefont
  {K.}~\bibnamefont {Spirin}}, \bibinfo {author} {\bibfnamefont
  {V.}~\bibnamefont {Aleshkin}}, \bibinfo {author} {\bibfnamefont
  {V.}~\bibnamefont {Gavrilenko}}, \ and\ \bibinfo {author} {\bibfnamefont
  {W.}~\bibnamefont {Knap}},\ }\href {\doibase 10.1103/PhysRevB.86.205420}
  {\bibfield  {journal} {\bibinfo  {journal} {Phys. Rev. B}\ }\textbf {\bibinfo
  {volume} {86}},\ \bibinfo {pages} {205420} (\bibinfo {year}
  {2012})}\BibitemShut {NoStop}%
\bibitem [{\citenamefont {Ludwig}\ \emph {et~al.}(2014)\citenamefont {Ludwig},
  \citenamefont {Vasilyev}, \citenamefont {Mikhailov}, \citenamefont
  {Poumirol}, \citenamefont {Jiang}, \citenamefont {Vafek},\ and\ \citenamefont
  {Smirnov}}]{Q6c}%
  \BibitemOpen
  \bibfield  {author} {\bibinfo {author} {\bibfnamefont {J.}~\bibnamefont
  {Ludwig}}, \bibinfo {author} {\bibfnamefont {Y.~B.}\ \bibnamefont
  {Vasilyev}}, \bibinfo {author} {\bibfnamefont {N.~N.}\ \bibnamefont
  {Mikhailov}}, \bibinfo {author} {\bibfnamefont {J.~M.}\ \bibnamefont
  {Poumirol}}, \bibinfo {author} {\bibfnamefont {Z.}~\bibnamefont {Jiang}},
  \bibinfo {author} {\bibfnamefont {O.}~\bibnamefont {Vafek}}, \ and\ \bibinfo
  {author} {\bibfnamefont {D.}~\bibnamefont {Smirnov}},\ }\href {\doibase
  10.1103/PhysRevB.89.241406} {\bibfield  {journal} {\bibinfo  {journal} {Phys.
  Rev. B}\ }\textbf {\bibinfo {volume} {89}},\ \bibinfo {pages} {241406}
  (\bibinfo {year} {2014})}\BibitemShut {NoStop}%
\bibitem [{\citenamefont {Ikonnikov}\ \emph {et~al.}(2016)\citenamefont
  {Ikonnikov}, \citenamefont {Krishtopenko}, \citenamefont {Drachenko},
  \citenamefont {Goiran}, \citenamefont {Zholudev}, \citenamefont {Platonov},
  \citenamefont {Kudasov}, \citenamefont {Korshunov}, \citenamefont {Maslov},
  \citenamefont {Makarov}, \citenamefont {Surdin}, \citenamefont {Philippov},
  \citenamefont {Marcinkiewicz}, \citenamefont {Ruffenach}, \citenamefont
  {Teppe}, \citenamefont {Knap}, \citenamefont {Mikhailov}, \citenamefont
  {Dvoretsky},\ and\ \citenamefont {Gavrilenko}}]{Q6d}%
  \BibitemOpen
  \bibfield  {author} {\bibinfo {author} {\bibfnamefont {A.~V.}\ \bibnamefont
  {Ikonnikov}}, \bibinfo {author} {\bibfnamefont {S.~S.}\ \bibnamefont
  {Krishtopenko}}, \bibinfo {author} {\bibfnamefont {O.}~\bibnamefont
  {Drachenko}}, \bibinfo {author} {\bibfnamefont {M.}~\bibnamefont {Goiran}},
  \bibinfo {author} {\bibfnamefont {M.~S.}\ \bibnamefont {Zholudev}}, \bibinfo
  {author} {\bibfnamefont {V.~V.}\ \bibnamefont {Platonov}}, \bibinfo {author}
  {\bibfnamefont {Y.~B.}\ \bibnamefont {Kudasov}}, \bibinfo {author}
  {\bibfnamefont {A.~S.}\ \bibnamefont {Korshunov}}, \bibinfo {author}
  {\bibfnamefont {D.~A.}\ \bibnamefont {Maslov}}, \bibinfo {author}
  {\bibfnamefont {I.~V.}\ \bibnamefont {Makarov}}, \bibinfo {author}
  {\bibfnamefont {O.~M.}\ \bibnamefont {Surdin}}, \bibinfo {author}
  {\bibfnamefont {A.~V.}\ \bibnamefont {Philippov}}, \bibinfo {author}
  {\bibfnamefont {M.}~\bibnamefont {Marcinkiewicz}}, \bibinfo {author}
  {\bibfnamefont {S.}~\bibnamefont {Ruffenach}}, \bibinfo {author}
  {\bibfnamefont {F.}~\bibnamefont {Teppe}}, \bibinfo {author} {\bibfnamefont
  {W.}~\bibnamefont {Knap}}, \bibinfo {author} {\bibfnamefont {N.~N.}\
  \bibnamefont {Mikhailov}}, \bibinfo {author} {\bibfnamefont {S.~A.}\
  \bibnamefont {Dvoretsky}}, \ and\ \bibinfo {author} {\bibfnamefont {V.~I.}\
  \bibnamefont {Gavrilenko}},\ }\href {\doibase 10.1103/PhysRevB.94.155421}
  {\bibfield  {journal} {\bibinfo  {journal} {Phys. Rev. B}\ }\textbf {\bibinfo
  {volume} {94}},\ \bibinfo {pages} {155421} (\bibinfo {year}
  {2016})}\BibitemShut {NoStop}%
\bibitem [{\citenamefont {Kotov}\ \emph {et~al.}(2012)\citenamefont {Kotov},
  \citenamefont {Uchoa}, \citenamefont {Pereira}, \citenamefont {Guinea},\ and\
  \citenamefont {Castro~Neto}}]{Q13}%
  \BibitemOpen
  \bibfield  {author} {\bibinfo {author} {\bibfnamefont {V.~N.}\ \bibnamefont
  {Kotov}}, \bibinfo {author} {\bibfnamefont {B.}~\bibnamefont {Uchoa}},
  \bibinfo {author} {\bibfnamefont {V.~M.}\ \bibnamefont {Pereira}}, \bibinfo
  {author} {\bibfnamefont {F.}~\bibnamefont {Guinea}}, \ and\ \bibinfo {author}
  {\bibfnamefont {A.~H.}\ \bibnamefont {Castro~Neto}},\ }\href {\doibase
  10.1103/RevModPhys.84.1067} {\bibfield  {journal} {\bibinfo  {journal} {Rev.
  Mod. Phys.}\ }\textbf {\bibinfo {volume} {84}},\ \bibinfo {pages} {1067}
  (\bibinfo {year} {2012})}\BibitemShut {NoStop}%
\bibitem [{\citenamefont {Wiedmann}\ \emph {et~al.}(2015)\citenamefont
  {Wiedmann}, \citenamefont {Jost}, \citenamefont {Thienel}, \citenamefont
  {Br\"une}, \citenamefont {Leubner}, \citenamefont {Buhmann}, \citenamefont
  {Molenkamp}, \citenamefont {Maan},\ and\ \citenamefont {Zeitler}}]{Q7}%
  \BibitemOpen
  \bibfield  {author} {\bibinfo {author} {\bibfnamefont {S.}~\bibnamefont
  {Wiedmann}}, \bibinfo {author} {\bibfnamefont {A.}~\bibnamefont {Jost}},
  \bibinfo {author} {\bibfnamefont {C.}~\bibnamefont {Thienel}}, \bibinfo
  {author} {\bibfnamefont {C.}~\bibnamefont {Br\"une}}, \bibinfo {author}
  {\bibfnamefont {P.}~\bibnamefont {Leubner}}, \bibinfo {author} {\bibfnamefont
  {H.}~\bibnamefont {Buhmann}}, \bibinfo {author} {\bibfnamefont {L.~W.}\
  \bibnamefont {Molenkamp}}, \bibinfo {author} {\bibfnamefont {J.~C.}\
  \bibnamefont {Maan}}, \ and\ \bibinfo {author} {\bibfnamefont
  {U.}~\bibnamefont {Zeitler}},\ }\href {\doibase 10.1103/PhysRevB.91.205311}
  {\bibfield  {journal} {\bibinfo  {journal} {Phys. Rev. B}\ }\textbf {\bibinfo
  {volume} {91}},\ \bibinfo {pages} {205311} (\bibinfo {year}
  {2015})}\BibitemShut {NoStop}%
\bibitem [{\citenamefont {Krishtopenko}\ \emph
  {et~al.}(2016{\natexlab{a}})\citenamefont {Krishtopenko}, \citenamefont
  {Yahniuk}, \citenamefont {But}, \citenamefont {Gavrilenko}, \citenamefont
  {Knap},\ and\ \citenamefont {Teppe}}]{Q8}%
  \BibitemOpen
  \bibfield  {author} {\bibinfo {author} {\bibfnamefont {S.~S.}\ \bibnamefont
  {Krishtopenko}}, \bibinfo {author} {\bibfnamefont {I.}~\bibnamefont
  {Yahniuk}}, \bibinfo {author} {\bibfnamefont {D.~B.}\ \bibnamefont {But}},
  \bibinfo {author} {\bibfnamefont {V.~I.}\ \bibnamefont {Gavrilenko}},
  \bibinfo {author} {\bibfnamefont {W.}~\bibnamefont {Knap}}, \ and\ \bibinfo
  {author} {\bibfnamefont {F.}~\bibnamefont {Teppe}},\ }\href {\doibase
  10.1103/PhysRevB.94.245402} {\bibfield  {journal} {\bibinfo  {journal} {Phys.
  Rev. B}\ }\textbf {\bibinfo {volume} {94}},\ \bibinfo {pages} {245402}
  (\bibinfo {year} {2016}{\natexlab{a}})}\BibitemShut {NoStop}%
\bibitem [{\citenamefont {Marcinkiewicz}\ \emph {et~al.}(2017)\citenamefont
  {Marcinkiewicz}, \citenamefont {Ruffenach}, \citenamefont {Krishtopenko},
  \citenamefont {Kadykov}, \citenamefont {Consejo}, \citenamefont {But},
  \citenamefont {Desrat}, \citenamefont {Knap}, \citenamefont {Torres},
  \citenamefont {Ikonnikov}, \citenamefont {Spirin}, \citenamefont {Morozov},
  \citenamefont {Gavrilenko}, \citenamefont {Mikhailov}, \citenamefont
  {Dvoretskii},\ and\ \citenamefont {Teppe}}]{Q9}%
  \BibitemOpen
  \bibfield  {author} {\bibinfo {author} {\bibfnamefont {M.}~\bibnamefont
  {Marcinkiewicz}}, \bibinfo {author} {\bibfnamefont {S.}~\bibnamefont
  {Ruffenach}}, \bibinfo {author} {\bibfnamefont {S.~S.}\ \bibnamefont
  {Krishtopenko}}, \bibinfo {author} {\bibfnamefont {A.~M.}\ \bibnamefont
  {Kadykov}}, \bibinfo {author} {\bibfnamefont {C.}~\bibnamefont {Consejo}},
  \bibinfo {author} {\bibfnamefont {D.~B.}\ \bibnamefont {But}}, \bibinfo
  {author} {\bibfnamefont {W.}~\bibnamefont {Desrat}}, \bibinfo {author}
  {\bibfnamefont {W.}~\bibnamefont {Knap}}, \bibinfo {author} {\bibfnamefont
  {J.}~\bibnamefont {Torres}}, \bibinfo {author} {\bibfnamefont {A.~V.}\
  \bibnamefont {Ikonnikov}}, \bibinfo {author} {\bibfnamefont {K.~E.}\
  \bibnamefont {Spirin}}, \bibinfo {author} {\bibfnamefont {S.~V.}\
  \bibnamefont {Morozov}}, \bibinfo {author} {\bibfnamefont {V.~I.}\
  \bibnamefont {Gavrilenko}}, \bibinfo {author} {\bibfnamefont {N.~N.}\
  \bibnamefont {Mikhailov}}, \bibinfo {author} {\bibfnamefont {S.~A.}\
  \bibnamefont {Dvoretskii}}, \ and\ \bibinfo {author} {\bibfnamefont
  {F.}~\bibnamefont {Teppe}},\ }\href {\doibase 10.1103/PhysRevB.96.035405}
  {\bibfield  {journal} {\bibinfo  {journal} {Phys. Rev. B}\ }\textbf {\bibinfo
  {volume} {96}},\ \bibinfo {pages} {035405} (\bibinfo {year}
  {2017})}\BibitemShut {NoStop}%
\bibitem [{\citenamefont {Kadykov}\ \emph {et~al.}(2018)\citenamefont
  {Kadykov}, \citenamefont {Krishtopenko}, \citenamefont {Jouault},
  \citenamefont {Desrat}, \citenamefont {Knap}, \citenamefont {Ruffenach},
  \citenamefont {Consejo}, \citenamefont {Torres}, \citenamefont {Morozov},
  \citenamefont {Mikhailov}, \citenamefont {Dvoretskii},\ and\ \citenamefont
  {Teppe}}]{Q10}%
  \BibitemOpen
  \bibfield  {author} {\bibinfo {author} {\bibfnamefont {A.~M.}\ \bibnamefont
  {Kadykov}}, \bibinfo {author} {\bibfnamefont {S.~S.}\ \bibnamefont
  {Krishtopenko}}, \bibinfo {author} {\bibfnamefont {B.}~\bibnamefont
  {Jouault}}, \bibinfo {author} {\bibfnamefont {W.}~\bibnamefont {Desrat}},
  \bibinfo {author} {\bibfnamefont {W.}~\bibnamefont {Knap}}, \bibinfo {author}
  {\bibfnamefont {S.}~\bibnamefont {Ruffenach}}, \bibinfo {author}
  {\bibfnamefont {C.}~\bibnamefont {Consejo}}, \bibinfo {author} {\bibfnamefont
  {J.}~\bibnamefont {Torres}}, \bibinfo {author} {\bibfnamefont {S.~V.}\
  \bibnamefont {Morozov}}, \bibinfo {author} {\bibfnamefont {N.~N.}\
  \bibnamefont {Mikhailov}}, \bibinfo {author} {\bibfnamefont {S.~A.}\
  \bibnamefont {Dvoretskii}}, \ and\ \bibinfo {author} {\bibfnamefont
  {F.}~\bibnamefont {Teppe}},\ }\href {\doibase 10.1103/PhysRevLett.120.086401}
  {\bibfield  {journal} {\bibinfo  {journal} {Phys. Rev. Lett.}\ }\textbf
  {\bibinfo {volume} {120}},\ \bibinfo {pages} {086401} (\bibinfo {year}
  {2018})}\BibitemShut {NoStop}%
\bibitem [{\citenamefont {Teppe}\ \emph {et~al.}(2016)\citenamefont {Teppe},
  \citenamefont {Marcinkiewicz}, \citenamefont {Krishtopenko}, \citenamefont
  {Ruffenach}, \citenamefont {Consejo}, \citenamefont {Kadykov}, \citenamefont
  {Desrat}, \citenamefont {But}, \citenamefont {Knap}, \citenamefont {Ludwig},
  \citenamefont {Moon}, \citenamefont {Smirnov}, \citenamefont {Orlita},
  \citenamefont {Jiang}, \citenamefont {Morozov}, \citenamefont {Gavrilenko},
  \citenamefont {Mikhailov},\ and\ \citenamefont {Dvoretskii}}]{Q11}%
  \BibitemOpen
  \bibfield  {author} {\bibinfo {author} {\bibfnamefont {F.}~\bibnamefont
  {Teppe}}, \bibinfo {author} {\bibfnamefont {M.}~\bibnamefont
  {Marcinkiewicz}}, \bibinfo {author} {\bibfnamefont {S.~S.}\ \bibnamefont
  {Krishtopenko}}, \bibinfo {author} {\bibfnamefont {S.}~\bibnamefont
  {Ruffenach}}, \bibinfo {author} {\bibfnamefont {C.}~\bibnamefont {Consejo}},
  \bibinfo {author} {\bibfnamefont {A.~M.}\ \bibnamefont {Kadykov}}, \bibinfo
  {author} {\bibfnamefont {W.}~\bibnamefont {Desrat}}, \bibinfo {author}
  {\bibfnamefont {D.}~\bibnamefont {But}}, \bibinfo {author} {\bibfnamefont
  {W.}~\bibnamefont {Knap}}, \bibinfo {author} {\bibfnamefont {J.}~\bibnamefont
  {Ludwig}}, \bibinfo {author} {\bibfnamefont {S.}~\bibnamefont {Moon}},
  \bibinfo {author} {\bibfnamefont {D.}~\bibnamefont {Smirnov}}, \bibinfo
  {author} {\bibfnamefont {M.}~\bibnamefont {Orlita}}, \bibinfo {author}
  {\bibfnamefont {Z.}~\bibnamefont {Jiang}}, \bibinfo {author} {\bibfnamefont
  {S.~V.}\ \bibnamefont {Morozov}}, \bibinfo {author} {\bibfnamefont
  {V.}~\bibnamefont {Gavrilenko}}, \bibinfo {author} {\bibfnamefont {N.~N.}\
  \bibnamefont {Mikhailov}}, \ and\ \bibinfo {author} {\bibfnamefont {S.~A.}\
  \bibnamefont {Dvoretskii}},\ }\href {\doibase 10.1038/ncomms12576} {\bibfield
   {journal} {\bibinfo  {journal} {Nat. Commun.}\ }\textbf {\bibinfo {volume}
  {7}},\ \bibinfo {pages} {12576} (\bibinfo {year} {2016})}\BibitemShut
  {NoStop}%
\bibitem [{\citenamefont {Miao}\ \emph {et~al.}(2012)\citenamefont {Miao},
  \citenamefont {Yan}, \citenamefont {Van~de Walle}, \citenamefont {Lou},
  \citenamefont {Li},\ and\ \citenamefont {Chang}}]{Q14}%
  \BibitemOpen
  \bibfield  {author} {\bibinfo {author} {\bibfnamefont {M.~S.}\ \bibnamefont
  {Miao}}, \bibinfo {author} {\bibfnamefont {Q.}~\bibnamefont {Yan}}, \bibinfo
  {author} {\bibfnamefont {C.~G.}\ \bibnamefont {Van~de Walle}}, \bibinfo
  {author} {\bibfnamefont {W.~K.}\ \bibnamefont {Lou}}, \bibinfo {author}
  {\bibfnamefont {L.~L.}\ \bibnamefont {Li}}, \ and\ \bibinfo {author}
  {\bibfnamefont {K.}~\bibnamefont {Chang}},\ }\href {\doibase
  10.1103/PhysRevLett.109.186803} {\bibfield  {journal} {\bibinfo  {journal}
  {Phys. Rev. Lett.}\ }\textbf {\bibinfo {volume} {109}},\ \bibinfo {pages}
  {186803} (\bibinfo {year} {2012})}\BibitemShut {NoStop}%
\bibitem [{\citenamefont {Zhang}\ \emph {et~al.}(2013)\citenamefont {Zhang},
  \citenamefont {Lou}, \citenamefont {Miao}, \citenamefont {Zhang},\ and\
  \citenamefont {Chang}}]{Q15}%
  \BibitemOpen
  \bibfield  {author} {\bibinfo {author} {\bibfnamefont {D.}~\bibnamefont
  {Zhang}}, \bibinfo {author} {\bibfnamefont {W.}~\bibnamefont {Lou}}, \bibinfo
  {author} {\bibfnamefont {M.}~\bibnamefont {Miao}}, \bibinfo {author}
  {\bibfnamefont {S.-C.}\ \bibnamefont {Zhang}}, \ and\ \bibinfo {author}
  {\bibfnamefont {K.}~\bibnamefont {Chang}},\ }\href {\doibase
  10.1103/PhysRevLett.111.156402} {\bibfield  {journal} {\bibinfo  {journal}
  {Phys. Rev. Lett.}\ }\textbf {\bibinfo {volume} {111}},\ \bibinfo {pages}
  {156402} (\bibinfo {year} {2013})}\BibitemShut {NoStop}%
\bibitem [{\citenamefont {Krishtopenko}\ and\ \citenamefont
  {Teppe}(2018{\natexlab{a}})}]{Q16}%
  \BibitemOpen
  \bibfield  {author} {\bibinfo {author} {\bibfnamefont {S.~S.}\ \bibnamefont
  {Krishtopenko}}\ and\ \bibinfo {author} {\bibfnamefont {F.}~\bibnamefont
  {Teppe}},\ }\href {\doibase 10.1126/sciadv.aap7529} {\bibfield  {journal}
  {\bibinfo  {journal} {Sci. Adv.}\ }\textbf {\bibinfo {volume} {4}},\ \bibinfo
  {pages} {eaap7529} (\bibinfo {year} {2018}{\natexlab{a}})}\BibitemShut
  {NoStop}%
\bibitem [{\citenamefont {Krishtopenko}\ \emph {et~al.}(2018)\citenamefont
  {Krishtopenko}, \citenamefont {Ruffenach}, \citenamefont {Gonzalez-Posada},
  \citenamefont {Boissier}, \citenamefont {Marcinkiewicz}, \citenamefont
  {Fadeev}, \citenamefont {Kadykov}, \citenamefont {Rumyantsev}, \citenamefont
  {Morozov}, \citenamefont {Gavrilenko}, \citenamefont {Consejo}, \citenamefont
  {Desrat}, \citenamefont {Jouault}, \citenamefont {Knap}, \citenamefont
  {Tourni\'e},\ and\ \citenamefont {Teppe}}]{Q17}%
  \BibitemOpen
  \bibfield  {author} {\bibinfo {author} {\bibfnamefont {S.~S.}\ \bibnamefont
  {Krishtopenko}}, \bibinfo {author} {\bibfnamefont {S.}~\bibnamefont
  {Ruffenach}}, \bibinfo {author} {\bibfnamefont {F.}~\bibnamefont
  {Gonzalez-Posada}}, \bibinfo {author} {\bibfnamefont {G.}~\bibnamefont
  {Boissier}}, \bibinfo {author} {\bibfnamefont {M.}~\bibnamefont
  {Marcinkiewicz}}, \bibinfo {author} {\bibfnamefont {M.~A.}\ \bibnamefont
  {Fadeev}}, \bibinfo {author} {\bibfnamefont {A.~M.}\ \bibnamefont {Kadykov}},
  \bibinfo {author} {\bibfnamefont {V.~V.}\ \bibnamefont {Rumyantsev}},
  \bibinfo {author} {\bibfnamefont {S.~V.}\ \bibnamefont {Morozov}}, \bibinfo
  {author} {\bibfnamefont {V.~I.}\ \bibnamefont {Gavrilenko}}, \bibinfo
  {author} {\bibfnamefont {C.}~\bibnamefont {Consejo}}, \bibinfo {author}
  {\bibfnamefont {W.}~\bibnamefont {Desrat}}, \bibinfo {author} {\bibfnamefont
  {B.}~\bibnamefont {Jouault}}, \bibinfo {author} {\bibfnamefont
  {W.}~\bibnamefont {Knap}}, \bibinfo {author} {\bibfnamefont {E.}~\bibnamefont
  {Tourni\'e}}, \ and\ \bibinfo {author} {\bibfnamefont {F.}~\bibnamefont
  {Teppe}},\ }\href {\doibase 10.1103/PhysRevB.97.245419} {\bibfield  {journal}
  {\bibinfo  {journal} {Phys. Rev. B}\ }\textbf {\bibinfo {volume} {97}},\
  \bibinfo {pages} {245419} (\bibinfo {year} {2018})}\BibitemShut {NoStop}%
\bibitem [{\citenamefont {Pikulin}\ and\ \citenamefont {Hyart}(2014)}]{Q19}%
  \BibitemOpen
  \bibfield  {author} {\bibinfo {author} {\bibfnamefont {D.~I.}\ \bibnamefont
  {Pikulin}}\ and\ \bibinfo {author} {\bibfnamefont {T.}~\bibnamefont
  {Hyart}},\ }\href {\doibase 10.1103/PhysRevLett.112.176403} {\bibfield
  {journal} {\bibinfo  {journal} {Phys. Rev. Lett.}\ }\textbf {\bibinfo
  {volume} {112}},\ \bibinfo {pages} {176403} (\bibinfo {year}
  {2014})}\BibitemShut {NoStop}%
\bibitem [{\citenamefont {Budich}\ \emph {et~al.}(2014)\citenamefont {Budich},
  \citenamefont {Trauzettel},\ and\ \citenamefont {Michetti}}]{Q20}%
  \BibitemOpen
  \bibfield  {author} {\bibinfo {author} {\bibfnamefont {J.~C.}\ \bibnamefont
  {Budich}}, \bibinfo {author} {\bibfnamefont {B.}~\bibnamefont {Trauzettel}},
  \ and\ \bibinfo {author} {\bibfnamefont {P.}~\bibnamefont {Michetti}},\
  }\href {\doibase 10.1103/PhysRevLett.112.146405} {\bibfield  {journal}
  {\bibinfo  {journal} {Phys. Rev. Lett.}\ }\textbf {\bibinfo {volume} {112}},\
  \bibinfo {pages} {146405} (\bibinfo {year} {2014})}\BibitemShut {NoStop}%
\bibitem [{\citenamefont {Du}\ \emph {et~al.}(2017)\citenamefont {Du},
  \citenamefont {Li}, \citenamefont {Lou}, \citenamefont {Sullivan},
  \citenamefont {Chang}, \citenamefont {Kono},\ and\ \citenamefont {Du}}]{Q21}%
  \BibitemOpen
  \bibfield  {author} {\bibinfo {author} {\bibfnamefont {L.}~\bibnamefont
  {Du}}, \bibinfo {author} {\bibfnamefont {X.}~\bibnamefont {Li}}, \bibinfo
  {author} {\bibfnamefont {W.}~\bibnamefont {Lou}}, \bibinfo {author}
  {\bibfnamefont {G.}~\bibnamefont {Sullivan}}, \bibinfo {author}
  {\bibfnamefont {K.}~\bibnamefont {Chang}}, \bibinfo {author} {\bibfnamefont
  {J.}~\bibnamefont {Kono}}, \ and\ \bibinfo {author} {\bibfnamefont {R.-R.}\
  \bibnamefont {Du}},\ }\href@noop {} {\bibfield  {journal} {\bibinfo
  {journal} {Nat. Commun.}\ }\textbf {\bibinfo {volume} {8}},\ \bibinfo {pages}
  {1971} (\bibinfo {year} {2017})}\BibitemShut {NoStop}%
\bibitem [{\citenamefont {Xue}\ and\ \citenamefont {MacDonald}(2018)}]{Q22}%
  \BibitemOpen
  \bibfield  {author} {\bibinfo {author} {\bibfnamefont {F.}~\bibnamefont
  {Xue}}\ and\ \bibinfo {author} {\bibfnamefont {A.~H.}\ \bibnamefont
  {MacDonald}},\ }\href {\doibase 10.1103/PhysRevLett.120.186802} {\bibfield
  {journal} {\bibinfo  {journal} {Phys. Rev. Lett.}\ }\textbf {\bibinfo
  {volume} {120}},\ \bibinfo {pages} {186802} (\bibinfo {year}
  {2018})}\BibitemShut {NoStop}%
\bibitem [{\citenamefont {Gauer}\ \emph {et~al.}(1993)\citenamefont {Gauer},
  \citenamefont {Scriba}, \citenamefont {Wixforth}, \citenamefont {Kotthaus},
  \citenamefont {Nguyen}, \citenamefont {Tuttle}, \citenamefont {English},\
  and\ \citenamefont {Kroemer}}]{Q23}%
  \BibitemOpen
  \bibfield  {author} {\bibinfo {author} {\bibfnamefont {C.}~\bibnamefont
  {Gauer}}, \bibinfo {author} {\bibfnamefont {J.}~\bibnamefont {Scriba}},
  \bibinfo {author} {\bibfnamefont {A.}~\bibnamefont {Wixforth}}, \bibinfo
  {author} {\bibfnamefont {J.~P.}\ \bibnamefont {Kotthaus}}, \bibinfo {author}
  {\bibfnamefont {C.}~\bibnamefont {Nguyen}}, \bibinfo {author} {\bibfnamefont
  {G.}~\bibnamefont {Tuttle}}, \bibinfo {author} {\bibfnamefont {J.~H.}\
  \bibnamefont {English}}, \ and\ \bibinfo {author} {\bibfnamefont
  {H.}~\bibnamefont {Kroemer}},\ }\href {\doibase 10.1088/0268-1242/8/1S/031}
  {\bibfield  {journal} {\bibinfo  {journal} {Semicond. Sci. Technol.}\
  }\textbf {\bibinfo {volume} {8}},\ \bibinfo {pages} {S137} (\bibinfo {year}
  {1993})}\BibitemShut {NoStop}%
\bibitem [{\citenamefont {Sadofyev}\ \emph {et~al.}(2005)\citenamefont
  {Sadofyev}, \citenamefont {Ramamoorthy}, \citenamefont {Bird}, \citenamefont
  {Johnson},\ and\ \citenamefont {Zhang}}]{Q24}%
  \BibitemOpen
  \bibfield  {author} {\bibinfo {author} {\bibfnamefont {Y.~G.}\ \bibnamefont
  {Sadofyev}}, \bibinfo {author} {\bibfnamefont {A.}~\bibnamefont
  {Ramamoorthy}}, \bibinfo {author} {\bibfnamefont {J.~P.}\ \bibnamefont
  {Bird}}, \bibinfo {author} {\bibfnamefont {S.~R.}\ \bibnamefont {Johnson}}, \
  and\ \bibinfo {author} {\bibfnamefont {Y.-H.}\ \bibnamefont {Zhang}},\ }\href
  {\doibase 10.1063/1.1926407} {\bibfield  {journal} {\bibinfo  {journal}
  {Appl. Phys. Lett.}\ }\textbf {\bibinfo {volume} {86}},\ \bibinfo {pages}
  {192109} (\bibinfo {year} {2005})}\BibitemShut {NoStop}%
\bibitem [{\citenamefont {Aleshkin}\ \emph
  {et~al.}(2005{\natexlab{a}})\citenamefont {Aleshkin}, \citenamefont
  {Gavrilenko}, \citenamefont {Gaponova}, \citenamefont {Ikonnikov},
  \citenamefont {Maremyanin}, \citenamefont {Morozov}, \citenamefont
  {Sadofyev}, \citenamefont {Johnson},\ and\ \citenamefont {Zhang}}]{Q25}%
  \BibitemOpen
  \bibfield  {author} {\bibinfo {author} {\bibfnamefont {V.~Y.}\ \bibnamefont
  {Aleshkin}}, \bibinfo {author} {\bibfnamefont {V.~I.}\ \bibnamefont
  {Gavrilenko}}, \bibinfo {author} {\bibfnamefont {D.~M.}\ \bibnamefont
  {Gaponova}}, \bibinfo {author} {\bibfnamefont {A.~V.}\ \bibnamefont
  {Ikonnikov}}, \bibinfo {author} {\bibfnamefont {K.~V.}\ \bibnamefont
  {Maremyanin}}, \bibinfo {author} {\bibfnamefont {S.~V.}\ \bibnamefont
  {Morozov}}, \bibinfo {author} {\bibfnamefont {Y.~G.}\ \bibnamefont
  {Sadofyev}}, \bibinfo {author} {\bibfnamefont {S.~R.}\ \bibnamefont
  {Johnson}}, \ and\ \bibinfo {author} {\bibfnamefont {Y.~H.}\ \bibnamefont
  {Zhang}},\ }\href {\doibase 10.1134/1.1852637} {\bibfield  {journal}
  {\bibinfo  {journal} {Semiconductors}\ }\textbf {\bibinfo {volume} {39}},\
  \bibinfo {pages} {22} (\bibinfo {year} {2005}{\natexlab{a}})}\BibitemShut
  {NoStop}%
\bibitem [{\citenamefont {Gavrilenko}\ \emph {et~al.}(2010)\citenamefont
  {Gavrilenko}, \citenamefont {Ikonnikov}, \citenamefont {Krishtopenko},
  \citenamefont {Lastovkin}, \citenamefont {Maremyanin}, \citenamefont
  {Sadofyev},\ and\ \citenamefont {Spirin}}]{Q26}%
  \BibitemOpen
  \bibfield  {author} {\bibinfo {author} {\bibfnamefont {V.~I.}\ \bibnamefont
  {Gavrilenko}}, \bibinfo {author} {\bibfnamefont {A.~V.}\ \bibnamefont
  {Ikonnikov}}, \bibinfo {author} {\bibfnamefont {S.~S.}\ \bibnamefont
  {Krishtopenko}}, \bibinfo {author} {\bibfnamefont {A.~A.}\ \bibnamefont
  {Lastovkin}}, \bibinfo {author} {\bibfnamefont {K.~V.}\ \bibnamefont
  {Maremyanin}}, \bibinfo {author} {\bibfnamefont {Y.~G.}\ \bibnamefont
  {Sadofyev}}, \ and\ \bibinfo {author} {\bibfnamefont {K.~E.}\ \bibnamefont
  {Spirin}},\ }\href {\doibase 10.1134/S106378261005012X} {\bibfield  {journal}
  {\bibinfo  {journal} {Semiconductors}\ }\textbf {\bibinfo {volume} {44}},\
  \bibinfo {pages} {616} (\bibinfo {year} {2010})}\BibitemShut {NoStop}%
\bibitem [{\citenamefont {Spirin}\ \emph {et~al.}(2012)\citenamefont {Spirin},
  \citenamefont {Kalinin}, \citenamefont {Krishtopenko}, \citenamefont
  {Maremyanin}, \citenamefont {Gavrilenko},\ and\ \citenamefont
  {Sadofyev}}]{Q27}%
  \BibitemOpen
  \bibfield  {author} {\bibinfo {author} {\bibfnamefont {K.~E.}\ \bibnamefont
  {Spirin}}, \bibinfo {author} {\bibfnamefont {K.~P.}\ \bibnamefont {Kalinin}},
  \bibinfo {author} {\bibfnamefont {S.~S.}\ \bibnamefont {Krishtopenko}},
  \bibinfo {author} {\bibfnamefont {K.~V.}\ \bibnamefont {Maremyanin}},
  \bibinfo {author} {\bibfnamefont {V.~I.}\ \bibnamefont {Gavrilenko}}, \ and\
  \bibinfo {author} {\bibfnamefont {Y.~G.}\ \bibnamefont {Sadofyev}},\ }\href
  {\doibase 10.1134/S1063782612110206} {\bibfield  {journal} {\bibinfo
  {journal} {Semiconductors}\ }\textbf {\bibinfo {volume} {46}},\ \bibinfo
  {pages} {1396} (\bibinfo {year} {2012})}\BibitemShut {NoStop}%
\bibitem [{\citenamefont {Tong}\ \emph {et~al.}(2017)\citenamefont {Tong},
  \citenamefont {Han}, \citenamefont {Li}, \citenamefont {Zhang}, \citenamefont
  {Sullivan},\ and\ \citenamefont {Du}}]{Q28}%
  \BibitemOpen
  \bibfield  {author} {\bibinfo {author} {\bibfnamefont {B.}~\bibnamefont
  {Tong}}, \bibinfo {author} {\bibfnamefont {Z.}~\bibnamefont {Han}}, \bibinfo
  {author} {\bibfnamefont {T.}~\bibnamefont {Li}}, \bibinfo {author}
  {\bibfnamefont {C.}~\bibnamefont {Zhang}}, \bibinfo {author} {\bibfnamefont
  {G.}~\bibnamefont {Sullivan}}, \ and\ \bibinfo {author} {\bibfnamefont
  {R.-R.}\ \bibnamefont {Du}},\ }\href {\doibase 10.1063/1.4993894} {\bibfield
  {journal} {\bibinfo  {journal} {AIP Advances}\ }\textbf {\bibinfo {volume}
  {7}},\ \bibinfo {pages} {075211} (\bibinfo {year} {2017})}\BibitemShut
  {NoStop}%
\bibitem [{\citenamefont {Ruffenach}\ \emph {et~al.}(2017)\citenamefont
  {Ruffenach}, \citenamefont {Krishtopenko}, \citenamefont {Bovkun},
  \citenamefont {Ikonnikov}, \citenamefont {Marcinkiewicz}, \citenamefont
  {Consejo}, \citenamefont {Potemski}, \citenamefont {Piot}, \citenamefont
  {Orlita}, \citenamefont {Semyagin}, \citenamefont {Putyato}, \citenamefont
  {Emelyanov}, \citenamefont {Preobrazhenskii}, \citenamefont {Knap},
  \citenamefont {Gonzalez-Posada}, \citenamefont {Boissier}, \citenamefont
  {Tourni\'{e}}, \citenamefont {Teppe},\ and\ \citenamefont
  {Gavrilenko}}]{Q18}%
  \BibitemOpen
  \bibfield  {author} {\bibinfo {author} {\bibfnamefont {S.}~\bibnamefont
  {Ruffenach}}, \bibinfo {author} {\bibfnamefont {S.~S.}\ \bibnamefont
  {Krishtopenko}}, \bibinfo {author} {\bibfnamefont {L.~S.}\ \bibnamefont
  {Bovkun}}, \bibinfo {author} {\bibfnamefont {A.~V.}\ \bibnamefont
  {Ikonnikov}}, \bibinfo {author} {\bibfnamefont {M.}~\bibnamefont
  {Marcinkiewicz}}, \bibinfo {author} {\bibfnamefont {C.}~\bibnamefont
  {Consejo}}, \bibinfo {author} {\bibfnamefont {M.}~\bibnamefont {Potemski}},
  \bibinfo {author} {\bibfnamefont {B.}~\bibnamefont {Piot}}, \bibinfo {author}
  {\bibfnamefont {M.}~\bibnamefont {Orlita}}, \bibinfo {author} {\bibfnamefont
  {B.~R.}\ \bibnamefont {Semyagin}}, \bibinfo {author} {\bibfnamefont {M.~A.}\
  \bibnamefont {Putyato}}, \bibinfo {author} {\bibfnamefont {E.~A.}\
  \bibnamefont {Emelyanov}}, \bibinfo {author} {\bibfnamefont {V.~V.}\
  \bibnamefont {Preobrazhenskii}}, \bibinfo {author} {\bibfnamefont
  {W.}~\bibnamefont {Knap}}, \bibinfo {author} {\bibfnamefont {F.}~\bibnamefont
  {Gonzalez-Posada}}, \bibinfo {author} {\bibfnamefont {G.}~\bibnamefont
  {Boissier}}, \bibinfo {author} {\bibfnamefont {E.}~\bibnamefont
  {Tourni\'{e}}}, \bibinfo {author} {\bibfnamefont {F.}~\bibnamefont {Teppe}},
  \ and\ \bibinfo {author} {\bibfnamefont {V.~I.}\ \bibnamefont {Gavrilenko}},\
  }\href {\doibase 10.1134/S0021364017230102} {\bibfield  {journal} {\bibinfo
  {journal} {JETP Lett.}\ }\textbf {\bibinfo {volume} {106}},\ \bibinfo {pages}
  {727} (\bibinfo {year} {2017})}\BibitemShut {NoStop}%
\bibitem [{\citenamefont {Knebl}\ \emph {et~al.}(2018)\citenamefont {Knebl},
  \citenamefont {Pfeffer}, \citenamefont {Schmid}, \citenamefont {Kamp},
  \citenamefont {Bastard}, \citenamefont {Batke}, \citenamefont {Worschech},
  \citenamefont {Hartmann},\ and\ \citenamefont {H\"ofling}}]{Q28a}%
  \BibitemOpen
  \bibfield  {author} {\bibinfo {author} {\bibfnamefont {G.}~\bibnamefont
  {Knebl}}, \bibinfo {author} {\bibfnamefont {P.}~\bibnamefont {Pfeffer}},
  \bibinfo {author} {\bibfnamefont {S.}~\bibnamefont {Schmid}}, \bibinfo
  {author} {\bibfnamefont {M.}~\bibnamefont {Kamp}}, \bibinfo {author}
  {\bibfnamefont {G.}~\bibnamefont {Bastard}}, \bibinfo {author} {\bibfnamefont
  {E.}~\bibnamefont {Batke}}, \bibinfo {author} {\bibfnamefont
  {L.}~\bibnamefont {Worschech}}, \bibinfo {author} {\bibfnamefont
  {F.}~\bibnamefont {Hartmann}}, \ and\ \bibinfo {author} {\bibfnamefont
  {S.}~\bibnamefont {H\"ofling}},\ }\href {\doibase 10.1103/PhysRevB.98.041301}
  {\bibfield  {journal} {\bibinfo  {journal} {Phys. Rev. B}\ }\textbf {\bibinfo
  {volume} {98}},\ \bibinfo {pages} {041301} (\bibinfo {year}
  {2018})}\BibitemShut {NoStop}%
\bibitem [{\citenamefont {Murakami}\ \emph {et~al.}(2007)\citenamefont
  {Murakami}, \citenamefont {Iso}, \citenamefont {Avishai}, \citenamefont
  {Onoda},\ and\ \citenamefont {Nagaosa}}]{Q29}%
  \BibitemOpen
  \bibfield  {author} {\bibinfo {author} {\bibfnamefont {S.}~\bibnamefont
  {Murakami}}, \bibinfo {author} {\bibfnamefont {S.}~\bibnamefont {Iso}},
  \bibinfo {author} {\bibfnamefont {Y.}~\bibnamefont {Avishai}}, \bibinfo
  {author} {\bibfnamefont {M.}~\bibnamefont {Onoda}}, \ and\ \bibinfo {author}
  {\bibfnamefont {N.}~\bibnamefont {Nagaosa}},\ }\href {\doibase
  10.1103/PhysRevB.76.205304} {\bibfield  {journal} {\bibinfo  {journal} {Phys.
  Rev. B}\ }\textbf {\bibinfo {volume} {76}},\ \bibinfo {pages} {205304}
  (\bibinfo {year} {2007})}\BibitemShut {NoStop}%
\bibitem [{SM()}]{SM}%
  \BibitemOpen
  \href@noop {} {\bibinfo  {journal} {See Supplemental Materials, which also
  contain Refs.~[43--51], for a brief discussion of simplified Dirac-like model
  in magnetic fields and details of magnetotransport measurements. The spectral
  studies of persistent photoconductivity of our sample are also provided
  therein}\ }\BibitemShut {NoStop}%
\bibitem [{\citenamefont {K\"{o}nig}\ \emph {et~al.}(2007)\citenamefont
  {K\"{o}nig}, \citenamefont {Wiedmann}, \citenamefont {Br\"{u}ne},
  \citenamefont {Roth}, \citenamefont {Buhmann}, \citenamefont {Molenkamp},
  \citenamefont {Qi},\ and\ \citenamefont {Zhang}}]{Q30}%
  \BibitemOpen
\bibfield  {journal} {  }\bibfield  {author} {\bibinfo {author} {\bibfnamefont
  {M.}~\bibnamefont {K\"{o}nig}}, \bibinfo {author} {\bibfnamefont
  {S.}~\bibnamefont {Wiedmann}}, \bibinfo {author} {\bibfnamefont
  {C.}~\bibnamefont {Br\"{u}ne}}, \bibinfo {author} {\bibfnamefont
  {A.}~\bibnamefont {Roth}}, \bibinfo {author} {\bibfnamefont {H.}~\bibnamefont
  {Buhmann}}, \bibinfo {author} {\bibfnamefont {L.~W.}\ \bibnamefont
  {Molenkamp}}, \bibinfo {author} {\bibfnamefont {X.-L.}\ \bibnamefont {Qi}}, \
  and\ \bibinfo {author} {\bibfnamefont {S.-C.}\ \bibnamefont {Zhang}},\ }\href
  {\doibase 10.1126/science.1148047} {\bibfield  {journal} {\bibinfo  {journal}
  {Science}\ }\textbf {\bibinfo {volume} {318}},\ \bibinfo {pages} {766}
  (\bibinfo {year} {2007})}\BibitemShut {NoStop}%
\bibitem [{\citenamefont {Klitzing}\ \emph {et~al.}(1980)\citenamefont
  {Klitzing}, \citenamefont {Dorda},\ and\ \citenamefont {Pepper}}]{Q35}%
  \BibitemOpen
  \bibfield  {author} {\bibinfo {author} {\bibfnamefont {K.~v.}\ \bibnamefont
  {Klitzing}}, \bibinfo {author} {\bibfnamefont {G.}~\bibnamefont {Dorda}}, \
  and\ \bibinfo {author} {\bibfnamefont {M.}~\bibnamefont {Pepper}},\ }\href
  {\doibase 10.1103/PhysRevLett.45.494} {\bibfield  {journal} {\bibinfo
  {journal} {Phys. Rev. Lett.}\ }\textbf {\bibinfo {volume} {45}},\ \bibinfo
  {pages} {494} (\bibinfo {year} {1980})}\BibitemShut {NoStop}%
\bibitem [{\citenamefont {Aleshkin}\ \emph
  {et~al.}(2005{\natexlab{b}})\citenamefont {Aleshkin}, \citenamefont
  {Gavrilenko}, \citenamefont {Ikonnikov}, \citenamefont {Sadofyev},
  \citenamefont {Bird}, \citenamefont {Johnson},\ and\ \citenamefont
  {Zhang}}]{Q32}%
  \BibitemOpen
  \bibfield  {author} {\bibinfo {author} {\bibfnamefont {V.~Y.}\ \bibnamefont
  {Aleshkin}}, \bibinfo {author} {\bibfnamefont {V.~I.}\ \bibnamefont
  {Gavrilenko}}, \bibinfo {author} {\bibfnamefont {A.~V.}\ \bibnamefont
  {Ikonnikov}}, \bibinfo {author} {\bibfnamefont {Y.~G.}\ \bibnamefont
  {Sadofyev}}, \bibinfo {author} {\bibfnamefont {J.~P.}\ \bibnamefont {Bird}},
  \bibinfo {author} {\bibfnamefont {S.~R.}\ \bibnamefont {Johnson}}, \ and\
  \bibinfo {author} {\bibfnamefont {Y.~H.}\ \bibnamefont {Zhang}},\ }\href
  {\doibase 10.1134/1.1852647} {\bibfield  {journal} {\bibinfo  {journal}
  {Semiconductors}\ }\textbf {\bibinfo {volume} {39}},\ \bibinfo {pages} {62}
  (\bibinfo {year} {2005}{\natexlab{b}})}\BibitemShut {NoStop}%
\bibitem [{\citenamefont {Kalinin}\ \emph {et~al.}(2013)\citenamefont
  {Kalinin}, \citenamefont {Krishtopenko}, \citenamefont {Maremyanin},
  \citenamefont {Spirin}, \citenamefont {Gavrilenko}, \citenamefont {Biryukov},
  \citenamefont {Baidus},\ and\ \citenamefont {Zvonkov}}]{Q32a}%
  \BibitemOpen
  \bibfield  {author} {\bibinfo {author} {\bibfnamefont {K.~P.}\ \bibnamefont
  {Kalinin}}, \bibinfo {author} {\bibfnamefont {S.~S.}\ \bibnamefont
  {Krishtopenko}}, \bibinfo {author} {\bibfnamefont {K.~V.}\ \bibnamefont
  {Maremyanin}}, \bibinfo {author} {\bibfnamefont {K.~E.}\ \bibnamefont
  {Spirin}}, \bibinfo {author} {\bibfnamefont {V.~I.}\ \bibnamefont
  {Gavrilenko}}, \bibinfo {author} {\bibfnamefont {A.~A.}\ \bibnamefont
  {Biryukov}}, \bibinfo {author} {\bibfnamefont {N.~V.}\ \bibnamefont
  {Baidus}}, \ and\ \bibinfo {author} {\bibfnamefont {B.~N.}\ \bibnamefont
  {Zvonkov}},\ }\href {\doibase 10.1134/S1063782613110092} {\bibfield
  {journal} {\bibinfo  {journal} {Semiconductors}\ }\textbf {\bibinfo {volume}
  {47}},\ \bibinfo {pages} {1485} (\bibinfo {year} {2013})}\BibitemShut
  {NoStop}%
\bibitem [{\citenamefont {Ikonnikov}\ \emph {et~al.}(2010)\citenamefont
  {Ikonnikov}, \citenamefont {Antonov}, \citenamefont {Lastovkin},
  \citenamefont {Gavrilenko}, \citenamefont {Sadof'ev},\ and\ \citenamefont
  {Samal}}]{Q31}%
  \BibitemOpen
  \bibfield  {author} {\bibinfo {author} {\bibfnamefont {A.~V.}\ \bibnamefont
  {Ikonnikov}}, \bibinfo {author} {\bibfnamefont {A.~V.}\ \bibnamefont
  {Antonov}}, \bibinfo {author} {\bibfnamefont {A.~A.}\ \bibnamefont
  {Lastovkin}}, \bibinfo {author} {\bibfnamefont {V.~I.}\ \bibnamefont
  {Gavrilenko}}, \bibinfo {author} {\bibfnamefont {Y.~G.}\ \bibnamefont
  {Sadof'ev}}, \ and\ \bibinfo {author} {\bibfnamefont {N.}~\bibnamefont
  {Samal}},\ }\href {\doibase 10.1134/S1063782610110175} {\bibfield  {journal}
  {\bibinfo  {journal} {Semiconductors}\ }\textbf {\bibinfo {volume} {44}},\
  \bibinfo {pages} {1467} (\bibinfo {year} {2010})}\BibitemShut {NoStop}%
\bibitem [{\citenamefont {Krishtopenko}\ and\ \citenamefont
  {Teppe}(2018{\natexlab{b}})}]{Q33}%
  \BibitemOpen
  \bibfield  {author} {\bibinfo {author} {\bibfnamefont {S.~S.}\ \bibnamefont
  {Krishtopenko}}\ and\ \bibinfo {author} {\bibfnamefont {F.}~\bibnamefont
  {Teppe}},\ }\href {\doibase 10.1103/PhysRevB.97.165408} {\bibfield  {journal}
  {\bibinfo  {journal} {Phys. Rev. B}\ }\textbf {\bibinfo {volume} {97}},\
  \bibinfo {pages} {165408} (\bibinfo {year} {2018}{\natexlab{b}})}\BibitemShut
  {NoStop}%
\bibitem [{\citenamefont {Semenikhin}\ \emph {et~al.}(2007)\citenamefont
  {Semenikhin}, \citenamefont {Zakharova}, \citenamefont {Nilsson},\ and\
  \citenamefont {Chao}}]{sm2}%
  \BibitemOpen
  \bibfield  {author} {\bibinfo {author} {\bibfnamefont {I.}~\bibnamefont
  {Semenikhin}}, \bibinfo {author} {\bibfnamefont {A.}~\bibnamefont
  {Zakharova}}, \bibinfo {author} {\bibfnamefont {K.}~\bibnamefont {Nilsson}},
  \ and\ \bibinfo {author} {\bibfnamefont {K.~A.}\ \bibnamefont {Chao}},\
  }\href {\doibase 10.1103/PhysRevB.76.035335} {\bibfield  {journal} {\bibinfo
  {journal} {Phys. Rev. B}\ }\textbf {\bibinfo {volume} {76}},\ \bibinfo
  {pages} {035335} (\bibinfo {year} {2007})}\BibitemShut {NoStop}%
\bibitem [{\citenamefont {Semenikhin}\ \emph {et~al.}(2008)\citenamefont
  {Semenikhin}, \citenamefont {Zakharova},\ and\ \citenamefont {Chao}}]{sm3}%
  \BibitemOpen
  \bibfield  {author} {\bibinfo {author} {\bibfnamefont {I.}~\bibnamefont
  {Semenikhin}}, \bibinfo {author} {\bibfnamefont {A.}~\bibnamefont
  {Zakharova}}, \ and\ \bibinfo {author} {\bibfnamefont {K.~A.}\ \bibnamefont
  {Chao}},\ }\href {\doibase 10.1103/PhysRevB.77.113307} {\bibfield  {journal}
  {\bibinfo  {journal} {Phys. Rev. B}\ }\textbf {\bibinfo {volume} {77}},\
  \bibinfo {pages} {113307} (\bibinfo {year} {2008})}\BibitemShut {NoStop}%
\bibitem [{\citenamefont {Charpentier}\ \emph {et~al.}(2013)\citenamefont
  {Charpentier}, \citenamefont {F{\"a}lt}, \citenamefont {Reichl},
  \citenamefont {Nichele}, \citenamefont {Pal}, \citenamefont {Pietsch},
  \citenamefont {Ihn}, \citenamefont {Ensslin},\ and\ \citenamefont
  {Wegscheider}}]{sm9}%
  \BibitemOpen
  \bibfield  {author} {\bibinfo {author} {\bibfnamefont {C.}~\bibnamefont
  {Charpentier}}, \bibinfo {author} {\bibfnamefont {S.}~\bibnamefont
  {F{\"a}lt}}, \bibinfo {author} {\bibfnamefont {C.}~\bibnamefont {Reichl}},
  \bibinfo {author} {\bibfnamefont {F.}~\bibnamefont {Nichele}}, \bibinfo
  {author} {\bibfnamefont {A.~N.}\ \bibnamefont {Pal}}, \bibinfo {author}
  {\bibfnamefont {P.}~\bibnamefont {Pietsch}}, \bibinfo {author} {\bibfnamefont
  {T.}~\bibnamefont {Ihn}}, \bibinfo {author} {\bibfnamefont {K.}~\bibnamefont
  {Ensslin}}, \ and\ \bibinfo {author} {\bibfnamefont {W.}~\bibnamefont
  {Wegscheider}},\ }\href@noop {} {\bibfield  {journal} {\bibinfo  {journal}
  {Appl. Phys. Lett.}\ }\textbf {\bibinfo {volume} {103}},\ \bibinfo {pages}
  {112102} (\bibinfo {year} {2013})}\BibitemShut {NoStop}%
\bibitem [{\citenamefont {Bolotin}\ \emph {et~al.}(2008)\citenamefont
  {Bolotin}, \citenamefont {Sikes}, \citenamefont {Jiang}, \citenamefont
  {Klima}, \citenamefont {Fudenberg}, \citenamefont {Hone}, \citenamefont
  {Kim},\ and\ \citenamefont {Stormer}}]{sm10}%
  \BibitemOpen
  \bibfield  {author} {\bibinfo {author} {\bibfnamefont {K.}~\bibnamefont
  {Bolotin}}, \bibinfo {author} {\bibfnamefont {K.}~\bibnamefont {Sikes}},
  \bibinfo {author} {\bibfnamefont {Z.}~\bibnamefont {Jiang}}, \bibinfo
  {author} {\bibfnamefont {M.}~\bibnamefont {Klima}}, \bibinfo {author}
  {\bibfnamefont {G.}~\bibnamefont {Fudenberg}}, \bibinfo {author}
  {\bibfnamefont {J.}~\bibnamefont {Hone}}, \bibinfo {author} {\bibfnamefont
  {P.}~\bibnamefont {Kim}}, \ and\ \bibinfo {author} {\bibfnamefont
  {H.}~\bibnamefont {Stormer}},\ }\href {\doibase
  https://doi.org/10.1016/j.ssc.2008.02.024} {\bibfield  {journal} {\bibinfo
  {journal} {Solid State Commun.}\ }\textbf {\bibinfo {volume} {146}},\
  \bibinfo {pages} {351 } (\bibinfo {year} {2008})}\BibitemShut {NoStop}%
\bibitem [{\citenamefont {Gusev}\ \emph {et~al.}(2017)\citenamefont {Gusev},
  \citenamefont {Kozlov}, \citenamefont {Levin}, \citenamefont {Kvon},
  \citenamefont {Mikhailov},\ and\ \citenamefont {Dvoretsky}}]{sm11}%
  \BibitemOpen
  \bibfield  {author} {\bibinfo {author} {\bibfnamefont {G.~M.}\ \bibnamefont
  {Gusev}}, \bibinfo {author} {\bibfnamefont {D.~A.}\ \bibnamefont {Kozlov}},
  \bibinfo {author} {\bibfnamefont {A.~D.}\ \bibnamefont {Levin}}, \bibinfo
  {author} {\bibfnamefont {Z.~D.}\ \bibnamefont {Kvon}}, \bibinfo {author}
  {\bibfnamefont {N.~N.}\ \bibnamefont {Mikhailov}}, \ and\ \bibinfo {author}
  {\bibfnamefont {S.~A.}\ \bibnamefont {Dvoretsky}},\ }\href {\doibase
  10.1103/PhysRevB.96.045304} {\bibfield  {journal} {\bibinfo  {journal} {Phys.
  Rev. B}\ }\textbf {\bibinfo {volume} {96}},\ \bibinfo {pages} {045304}
  (\bibinfo {year} {2017})}\BibitemShut {NoStop}%
\bibitem [{\citenamefont {Krishtopenko}\ \emph
  {et~al.}(2016{\natexlab{b}})\citenamefont {Krishtopenko}, \citenamefont
  {Knap},\ and\ \citenamefont {Teppe}}]{sm12}%
  \BibitemOpen
  \bibfield  {author} {\bibinfo {author} {\bibfnamefont {S.~S.}\ \bibnamefont
  {Krishtopenko}}, \bibinfo {author} {\bibfnamefont {W.}~\bibnamefont {Knap}},
  \ and\ \bibinfo {author} {\bibfnamefont {F.}~\bibnamefont {Teppe}},\ }\href
  {\doibase 10.1038/srep30755} {\bibfield  {journal} {\bibinfo  {journal} {Sci.
  Rep.}\ }\textbf {\bibinfo {volume} {6}},\ \bibinfo {pages} {30755} (\bibinfo
  {year} {2016}{\natexlab{b}})}\BibitemShut {NoStop}%
\bibitem [{\citenamefont {Rothe}\ \emph {et~al.}(2010)\citenamefont {Rothe},
  \citenamefont {Reinthaler}, \citenamefont {Liu}, \citenamefont {Molenkamp},
  \citenamefont {Zhang},\ and\ \citenamefont {Hankiewicz}}]{sm13}%
  \BibitemOpen
  \bibfield  {author} {\bibinfo {author} {\bibfnamefont {D.~G.}\ \bibnamefont
  {Rothe}}, \bibinfo {author} {\bibfnamefont {R.~W.}\ \bibnamefont
  {Reinthaler}}, \bibinfo {author} {\bibfnamefont {C.-X.}\ \bibnamefont {Liu}},
  \bibinfo {author} {\bibfnamefont {L.~W.}\ \bibnamefont {Molenkamp}}, \bibinfo
  {author} {\bibfnamefont {S.-C.}\ \bibnamefont {Zhang}}, \ and\ \bibinfo
  {author} {\bibfnamefont {E.~M.}\ \bibnamefont {Hankiewicz}},\ }\href
  {http://stacks.iop.org/1367-2630/12/i=6/a=065012} {\bibfield  {journal}
  {\bibinfo  {journal} {New J. Phys.}\ }\textbf {\bibinfo {volume} {12}},\
  \bibinfo {pages} {065012} (\bibinfo {year} {2010})}\BibitemShut {NoStop}%
\bibitem [{\citenamefont {Tuttle}\ \emph {et~al.}(1989)\citenamefont {Tuttle},
  \citenamefont {Kroemer},\ and\ \citenamefont {English}}]{sm6}%
  \BibitemOpen
  \bibfield  {author} {\bibinfo {author} {\bibfnamefont {G.}~\bibnamefont
  {Tuttle}}, \bibinfo {author} {\bibfnamefont {H.}~\bibnamefont {Kroemer}}, \
  and\ \bibinfo {author} {\bibfnamefont {J.~H.}\ \bibnamefont {English}},\
  }\href {\doibase 10.1063/1.343167} {\bibfield  {journal} {\bibinfo  {journal}
  {J. Appl. Phys.}\ }\textbf {\bibinfo {volume} {65}},\ \bibinfo {pages} {5239}
  (\bibinfo {year} {1989})}\BibitemShut {NoStop}%
\bibitem [{\citenamefont {Tuttle}\ \emph {et~al.}(1990)\citenamefont {Tuttle},
  \citenamefont {Kroemer},\ and\ \citenamefont {English}}]{sm6a}%
  \BibitemOpen
  \bibfield  {author} {\bibinfo {author} {\bibfnamefont {G.}~\bibnamefont
  {Tuttle}}, \bibinfo {author} {\bibfnamefont {H.}~\bibnamefont {Kroemer}}, \
  and\ \bibinfo {author} {\bibfnamefont {J.~H.}\ \bibnamefont {English}},\
  }\href {\doibase 10.1063/1.345426} {\bibfield  {journal} {\bibinfo  {journal}
  {J. Appl. Phys.}\ }\textbf {\bibinfo {volume} {67}},\ \bibinfo {pages} {3032}
  (\bibinfo {year} {1990})}\BibitemShut {NoStop}%
\end{thebibliography}

\begin{thebibliography}{21}%
\makeatletter
\providecommand \@ifxundefined [1]{%
 \@ifx{#1\undefined}
}%
\providecommand \@ifnum [1]{%
 \ifnum #1\expandafter \@firstoftwo
 \else \expandafter \@secondoftwo
 \fi
}%
\providecommand \@ifx [1]{%
 \ifx #1\expandafter \@firstoftwo
 \else \expandafter \@secondoftwo
 \fi
}%
\providecommand \natexlab [1]{#1}%
\providecommand \enquote  [1]{``#1''}%
\providecommand \bibnamefont  [1]{#1}%
\providecommand \bibfnamefont [1]{#1}%
\providecommand \citenamefont [1]{#1}%
\providecommand \href@noop [0]{\@secondoftwo}%
\providecommand \href [0]{\begingroup \@sanitize@url \@href}%
\providecommand \@href[1]{\@@startlink{#1}\@@href}%
\providecommand \@@href[1]{\endgroup#1\@@endlink}%
\providecommand \@sanitize@url [0]{\catcode `\\12\catcode `\$12\catcode
  `\&12\catcode `\#12\catcode `\^12\catcode `\_12\catcode `\%12\relax}%
\providecommand \@@startlink[1]{}%
\providecommand \@@endlink[0]{}%
\providecommand \url  [0]{\begingroup\@sanitize@url \@url }%
\providecommand \@url [1]{\endgroup\@href {#1}{\urlprefix }}%
\providecommand \urlprefix  [0]{URL }%
\providecommand \Eprint [0]{\href }%
\providecommand \doibase [0]{http://dx.doi.org/}%
\providecommand \selectlanguage [0]{\@gobble}%
\providecommand \bibinfo  [0]{\@secondoftwo}%
\providecommand \bibfield  [0]{\@secondoftwo}%
\providecommand \translation [1]{[#1]}%
\providecommand \BibitemOpen [0]{}%
\providecommand \bibitemStop [0]{}%
\providecommand \bibitemNoStop [0]{.\EOS\space}%
\providecommand \EOS [0]{\spacefactor3000\relax}%
\providecommand \BibitemShut  [1]{\csname bibitem#1\endcsname}%
\let\auto@bib@innerbib\@empty
\bibitem [{\citenamefont {Bernevig}\ \emph {et~al.}(2006)\citenamefont
  {Bernevig}, \citenamefont {Hughes},\ and\ \citenamefont {Zhang}}]{sm1}%
  \BibitemOpen
  \bibfield  {author} {\bibinfo {author} {\bibfnamefont {B.~A.}\ \bibnamefont
  {Bernevig}}, \bibinfo {author} {\bibfnamefont {T.~L.}\ \bibnamefont
  {Hughes}}, \ and\ \bibinfo {author} {\bibfnamefont {S.-C.}\ \bibnamefont
  {Zhang}},\ }\href {\doibase 10.1126/science.1133734} {\bibfield  {journal}
  {\bibinfo  {journal} {Science}\ }\textbf {\bibinfo {volume} {314}},\ \bibinfo
  {pages} {1757} (\bibinfo {year} {2006})}\BibitemShut {NoStop}%
\bibitem [{\citenamefont {Krishtopenko}\ \emph
  {et~al.}(2016{\natexlab{a}})\citenamefont {Krishtopenko}, \citenamefont
  {Yahniuk}, \citenamefont {But}, \citenamefont {Gavrilenko}, \citenamefont
  {Knap},\ and\ \citenamefont {Teppe}}]{sm7}%
  \BibitemOpen
  \bibfield  {author} {\bibinfo {author} {\bibfnamefont {S.~S.}\ \bibnamefont
  {Krishtopenko}}, \bibinfo {author} {\bibfnamefont {I.}~\bibnamefont
  {Yahniuk}}, \bibinfo {author} {\bibfnamefont {D.~B.}\ \bibnamefont {But}},
  \bibinfo {author} {\bibfnamefont {V.~I.}\ \bibnamefont {Gavrilenko}},
  \bibinfo {author} {\bibfnamefont {W.}~\bibnamefont {Knap}}, \ and\ \bibinfo
  {author} {\bibfnamefont {F.}~\bibnamefont {Teppe}},\ }\href {\doibase
  10.1103/PhysRevB.94.245402} {\bibfield  {journal} {\bibinfo  {journal} {Phys.
  Rev. B}\ }\textbf {\bibinfo {volume} {94}},\ \bibinfo {pages} {245402}
  (\bibinfo {year} {2016}{\natexlab{a}})}\BibitemShut {NoStop}%
\bibitem [{\citenamefont {Krishtopenko}\ \emph {et~al.}(2018)\citenamefont
  {Krishtopenko}, \citenamefont {Ruffenach}, \citenamefont {Gonzalez-Posada},
  \citenamefont {Boissier}, \citenamefont {Marcinkiewicz}, \citenamefont
  {Fadeev}, \citenamefont {Kadykov}, \citenamefont {Rumyantsev}, \citenamefont
  {Morozov}, \citenamefont {Gavrilenko}, \citenamefont {Consejo}, \citenamefont
  {Desrat}, \citenamefont {Jouault}, \citenamefont {Knap}, \citenamefont
  {Tourni\'e},\ and\ \citenamefont {Teppe}}]{sm8}%
  \BibitemOpen
  \bibfield  {author} {\bibinfo {author} {\bibfnamefont {S.~S.}\ \bibnamefont
  {Krishtopenko}}, \bibinfo {author} {\bibfnamefont {S.}~\bibnamefont
  {Ruffenach}}, \bibinfo {author} {\bibfnamefont {F.}~\bibnamefont
  {Gonzalez-Posada}}, \bibinfo {author} {\bibfnamefont {G.}~\bibnamefont
  {Boissier}}, \bibinfo {author} {\bibfnamefont {M.}~\bibnamefont
  {Marcinkiewicz}}, \bibinfo {author} {\bibfnamefont {M.~A.}\ \bibnamefont
  {Fadeev}}, \bibinfo {author} {\bibfnamefont {A.~M.}\ \bibnamefont {Kadykov}},
  \bibinfo {author} {\bibfnamefont {V.~V.}\ \bibnamefont {Rumyantsev}},
  \bibinfo {author} {\bibfnamefont {S.~V.}\ \bibnamefont {Morozov}}, \bibinfo
  {author} {\bibfnamefont {V.~I.}\ \bibnamefont {Gavrilenko}}, \bibinfo
  {author} {\bibfnamefont {C.}~\bibnamefont {Consejo}}, \bibinfo {author}
  {\bibfnamefont {W.}~\bibnamefont {Desrat}}, \bibinfo {author} {\bibfnamefont
  {B.}~\bibnamefont {Jouault}}, \bibinfo {author} {\bibfnamefont
  {W.}~\bibnamefont {Knap}}, \bibinfo {author} {\bibfnamefont {E.}~\bibnamefont
  {Tourni\'e}}, \ and\ \bibinfo {author} {\bibfnamefont {F.}~\bibnamefont
  {Teppe}},\ }\href {\doibase 10.1103/PhysRevB.97.245419} {\bibfield  {journal}
  {\bibinfo  {journal} {Phys. Rev. B}\ }\textbf {\bibinfo {volume} {97}},\
  \bibinfo {pages} {245419} (\bibinfo {year} {2018})}\BibitemShut {NoStop}%
\bibitem [{\citenamefont {Semenikhin}\ \emph {et~al.}(2007)\citenamefont
  {Semenikhin}, \citenamefont {Zakharova}, \citenamefont {Nilsson},\ and\
  \citenamefont {Chao}}]{sm2}%
  \BibitemOpen
  \bibfield  {author} {\bibinfo {author} {\bibfnamefont {I.}~\bibnamefont
  {Semenikhin}}, \bibinfo {author} {\bibfnamefont {A.}~\bibnamefont
  {Zakharova}}, \bibinfo {author} {\bibfnamefont {K.}~\bibnamefont {Nilsson}},
  \ and\ \bibinfo {author} {\bibfnamefont {K.~A.}\ \bibnamefont {Chao}},\
  }\href {\doibase 10.1103/PhysRevB.76.035335} {\bibfield  {journal} {\bibinfo
  {journal} {Phys. Rev. B}\ }\textbf {\bibinfo {volume} {76}},\ \bibinfo
  {pages} {035335} (\bibinfo {year} {2007})}\BibitemShut {NoStop}%
\bibitem [{\citenamefont {Semenikhin}\ \emph {et~al.}(2008)\citenamefont
  {Semenikhin}, \citenamefont {Zakharova},\ and\ \citenamefont {Chao}}]{sm3}%
  \BibitemOpen
  \bibfield  {author} {\bibinfo {author} {\bibfnamefont {I.}~\bibnamefont
  {Semenikhin}}, \bibinfo {author} {\bibfnamefont {A.}~\bibnamefont
  {Zakharova}}, \ and\ \bibinfo {author} {\bibfnamefont {K.~A.}\ \bibnamefont
  {Chao}},\ }\href {\doibase 10.1103/PhysRevB.77.113307} {\bibfield  {journal}
  {\bibinfo  {journal} {Phys. Rev. B}\ }\textbf {\bibinfo {volume} {77}},\
  \bibinfo {pages} {113307} (\bibinfo {year} {2008})}\BibitemShut {NoStop}%
\bibitem [{\citenamefont {K\"{o}nig}\ \emph {et~al.}(2007)\citenamefont
  {K\"{o}nig}, \citenamefont {Wiedmann}, \citenamefont {Br\"{u}ne},
  \citenamefont {Roth}, \citenamefont {Buhmann}, \citenamefont {Molenkamp},
  \citenamefont {Qi},\ and\ \citenamefont {Zhang}}]{sm4}%
  \BibitemOpen
  \bibfield  {author} {\bibinfo {author} {\bibfnamefont {M.}~\bibnamefont
  {K\"{o}nig}}, \bibinfo {author} {\bibfnamefont {S.}~\bibnamefont {Wiedmann}},
  \bibinfo {author} {\bibfnamefont {C.}~\bibnamefont {Br\"{u}ne}}, \bibinfo
  {author} {\bibfnamefont {A.}~\bibnamefont {Roth}}, \bibinfo {author}
  {\bibfnamefont {H.}~\bibnamefont {Buhmann}}, \bibinfo {author} {\bibfnamefont
  {L.~W.}\ \bibnamefont {Molenkamp}}, \bibinfo {author} {\bibfnamefont {X.-L.}\
  \bibnamefont {Qi}}, \ and\ \bibinfo {author} {\bibfnamefont {S.-C.}\
  \bibnamefont {Zhang}},\ }\href {\doibase 10.1126/science.1148047} {\bibfield
  {journal} {\bibinfo  {journal} {Science}\ }\textbf {\bibinfo {volume}
  {318}},\ \bibinfo {pages} {766} (\bibinfo {year} {2007})}\BibitemShut
  {NoStop}%
\bibitem [{\citenamefont {Castro~Neto}\ \emph {et~al.}(2009)\citenamefont
  {Castro~Neto}, \citenamefont {Guinea}, \citenamefont {Peres}, \citenamefont
  {Novoselov},\ and\ \citenamefont {Geim}}]{sm5}%
  \BibitemOpen
  \bibfield  {author} {\bibinfo {author} {\bibfnamefont {A.~H.}\ \bibnamefont
  {Castro~Neto}}, \bibinfo {author} {\bibfnamefont {F.}~\bibnamefont {Guinea}},
  \bibinfo {author} {\bibfnamefont {N.~M.~R.}\ \bibnamefont {Peres}}, \bibinfo
  {author} {\bibfnamefont {K.~S.}\ \bibnamefont {Novoselov}}, \ and\ \bibinfo
  {author} {\bibfnamefont {A.~K.}\ \bibnamefont {Geim}},\ }\href {\doibase
  10.1103/RevModPhys.81.109} {\bibfield  {journal} {\bibinfo  {journal} {Rev.
  Mod. Phys.}\ }\textbf {\bibinfo {volume} {81}},\ \bibinfo {pages} {109}
  (\bibinfo {year} {2009})}\BibitemShut {NoStop}%
\bibitem [{\citenamefont {Tuttle}\ \emph {et~al.}(1990)\citenamefont {Tuttle},
  \citenamefont {Kroemer},\ and\ \citenamefont {English}}]{sm6a}%
  \BibitemOpen
  \bibfield  {author} {\bibinfo {author} {\bibfnamefont {G.}~\bibnamefont
  {Tuttle}}, \bibinfo {author} {\bibfnamefont {H.}~\bibnamefont {Kroemer}}, \
  and\ \bibinfo {author} {\bibfnamefont {J.~H.}\ \bibnamefont {English}},\
  }\href {\doibase 10.1063/1.345426} {\bibfield  {journal} {\bibinfo  {journal}
  {J. Appl. Phys.}\ }\textbf {\bibinfo {volume} {67}},\ \bibinfo {pages} {3032}
  (\bibinfo {year} {1990})}\BibitemShut {NoStop}%
\bibitem [{\citenamefont {B\"{u}ttner}\ \emph {et~al.}(2011)\citenamefont
  {B\"{u}ttner}, \citenamefont {Liu}, \citenamefont {Tkachov}, \citenamefont
  {Novik}, \citenamefont {Br\"{u}ne}, \citenamefont {Buhmann}, \citenamefont
  {Hankiewicz}, \citenamefont {Recher}, \citenamefont {Trauzettel},
  \citenamefont {Zhang},\ and\ \citenamefont {Molenkamp}}]{smQ6a}%
  \BibitemOpen
  \bibfield  {author} {\bibinfo {author} {\bibfnamefont {B.}~\bibnamefont
  {B\"{u}ttner}}, \bibinfo {author} {\bibfnamefont {C.}~\bibnamefont {Liu}},
  \bibinfo {author} {\bibfnamefont {G.}~\bibnamefont {Tkachov}}, \bibinfo
  {author} {\bibfnamefont {E.}~\bibnamefont {Novik}}, \bibinfo {author}
  {\bibfnamefont {C.}~\bibnamefont {Br\"{u}ne}}, \bibinfo {author}
  {\bibfnamefont {H.}~\bibnamefont {Buhmann}}, \bibinfo {author} {\bibfnamefont
  {E.}~\bibnamefont {Hankiewicz}}, \bibinfo {author} {\bibfnamefont
  {P.}~\bibnamefont {Recher}}, \bibinfo {author} {\bibfnamefont
  {B.}~\bibnamefont {Trauzettel}}, \bibinfo {author} {\bibfnamefont
  {S.}~\bibnamefont {Zhang}}, \ and\ \bibinfo {author} {\bibfnamefont
  {L.}~\bibnamefont {Molenkamp}},\ }\href {\doibase 10.1038/nphys1914}
  {\bibfield  {journal} {\bibinfo  {journal} {Nat. Phys.}\ }\textbf {\bibinfo
  {volume} {7}},\ \bibinfo {pages} {418} (\bibinfo {year} {2011})}\BibitemShut
  {NoStop}%
\bibitem [{\citenamefont {Bolotin}\ \emph {et~al.}(2008)\citenamefont
  {Bolotin}, \citenamefont {Sikes}, \citenamefont {Jiang}, \citenamefont
  {Klima}, \citenamefont {Fudenberg}, \citenamefont {Hone}, \citenamefont
  {Kim},\ and\ \citenamefont {Stormer}}]{sm10}%
  \BibitemOpen
  \bibfield  {author} {\bibinfo {author} {\bibfnamefont {K.}~\bibnamefont
  {Bolotin}}, \bibinfo {author} {\bibfnamefont {K.}~\bibnamefont {Sikes}},
  \bibinfo {author} {\bibfnamefont {Z.}~\bibnamefont {Jiang}}, \bibinfo
  {author} {\bibfnamefont {M.}~\bibnamefont {Klima}}, \bibinfo {author}
  {\bibfnamefont {G.}~\bibnamefont {Fudenberg}}, \bibinfo {author}
  {\bibfnamefont {J.}~\bibnamefont {Hone}}, \bibinfo {author} {\bibfnamefont
  {P.}~\bibnamefont {Kim}}, \ and\ \bibinfo {author} {\bibfnamefont
  {H.}~\bibnamefont {Stormer}},\ }\href {\doibase
  https://doi.org/10.1016/j.ssc.2008.02.024} {\bibfield  {journal} {\bibinfo
  {journal} {Solid State Commun.}\ }\textbf {\bibinfo {volume} {146}},\
  \bibinfo {pages} {351 } (\bibinfo {year} {2008})}\BibitemShut {NoStop}%
\bibitem [{\citenamefont {Gusev}\ \emph {et~al.}(2017)\citenamefont {Gusev},
  \citenamefont {Kozlov}, \citenamefont {Levin}, \citenamefont {Kvon},
  \citenamefont {Mikhailov},\ and\ \citenamefont {Dvoretsky}}]{sm11}%
  \BibitemOpen
  \bibfield  {author} {\bibinfo {author} {\bibfnamefont {G.~M.}\ \bibnamefont
  {Gusev}}, \bibinfo {author} {\bibfnamefont {D.~A.}\ \bibnamefont {Kozlov}},
  \bibinfo {author} {\bibfnamefont {A.~D.}\ \bibnamefont {Levin}}, \bibinfo
  {author} {\bibfnamefont {Z.~D.}\ \bibnamefont {Kvon}}, \bibinfo {author}
  {\bibfnamefont {N.~N.}\ \bibnamefont {Mikhailov}}, \ and\ \bibinfo {author}
  {\bibfnamefont {S.~A.}\ \bibnamefont {Dvoretsky}},\ }\href {\doibase
  10.1103/PhysRevB.96.045304} {\bibfield  {journal} {\bibinfo  {journal} {Phys.
  Rev. B}\ }\textbf {\bibinfo {volume} {96}},\ \bibinfo {pages} {045304}
  (\bibinfo {year} {2017})}\BibitemShut {NoStop}%
\bibitem [{\citenamefont {Charpentier}\ \emph {et~al.}(2013)\citenamefont
  {Charpentier}, \citenamefont {F{\"a}lt}, \citenamefont {Reichl},
  \citenamefont {Nichele}, \citenamefont {Pal}, \citenamefont {Pietsch},
  \citenamefont {Ihn}, \citenamefont {Ensslin},\ and\ \citenamefont
  {Wegscheider}}]{sm9}%
  \BibitemOpen
  \bibfield  {author} {\bibinfo {author} {\bibfnamefont {C.}~\bibnamefont
  {Charpentier}}, \bibinfo {author} {\bibfnamefont {S.}~\bibnamefont
  {F{\"a}lt}}, \bibinfo {author} {\bibfnamefont {C.}~\bibnamefont {Reichl}},
  \bibinfo {author} {\bibfnamefont {F.}~\bibnamefont {Nichele}}, \bibinfo
  {author} {\bibfnamefont {A.~N.}\ \bibnamefont {Pal}}, \bibinfo {author}
  {\bibfnamefont {P.}~\bibnamefont {Pietsch}}, \bibinfo {author} {\bibfnamefont
  {T.}~\bibnamefont {Ihn}}, \bibinfo {author} {\bibfnamefont {K.}~\bibnamefont
  {Ensslin}}, \ and\ \bibinfo {author} {\bibfnamefont {W.}~\bibnamefont
  {Wegscheider}},\ }\href@noop {} {\bibfield  {journal} {\bibinfo  {journal}
  {Appl. Phys. Lett.}\ }\textbf {\bibinfo {volume} {103}},\ \bibinfo {pages}
  {112102} (\bibinfo {year} {2013})}\BibitemShut {NoStop}%
\bibitem [{\citenamefont {Krishtopenko}\ \emph
  {et~al.}(2016{\natexlab{b}})\citenamefont {Krishtopenko}, \citenamefont
  {Knap},\ and\ \citenamefont {Teppe}}]{sm12}%
  \BibitemOpen
  \bibfield  {author} {\bibinfo {author} {\bibfnamefont {S.~S.}\ \bibnamefont
  {Krishtopenko}}, \bibinfo {author} {\bibfnamefont {W.}~\bibnamefont {Knap}},
  \ and\ \bibinfo {author} {\bibfnamefont {F.}~\bibnamefont {Teppe}},\ }\href
  {\doibase 10.1038/srep30755} {\bibfield  {journal} {\bibinfo  {journal} {Sci.
  Rep.}\ }\textbf {\bibinfo {volume} {6}},\ \bibinfo {pages} {30755} (\bibinfo
  {year} {2016}{\natexlab{b}})}\BibitemShut {NoStop}%
\bibitem [{\citenamefont {Krishtopenko}\ and\ \citenamefont
  {Teppe}(2018)}]{smQ16}%
  \BibitemOpen
  \bibfield  {author} {\bibinfo {author} {\bibfnamefont {S.~S.}\ \bibnamefont
  {Krishtopenko}}\ and\ \bibinfo {author} {\bibfnamefont {F.}~\bibnamefont
  {Teppe}},\ }\href {\doibase 10.1126/sciadv.aap7529} {\bibfield  {journal}
  {\bibinfo  {journal} {Sci. Adv.}\ }\textbf {\bibinfo {volume} {4}},\ \bibinfo
  {pages} {eaap7529} (\bibinfo {year} {2018})}\BibitemShut {NoStop}%
\bibitem [{\citenamefont {Rothe}\ \emph {et~al.}(2010)\citenamefont {Rothe},
  \citenamefont {Reinthaler}, \citenamefont {Liu}, \citenamefont {Molenkamp},
  \citenamefont {Zhang},\ and\ \citenamefont {Hankiewicz}}]{sm13}%
  \BibitemOpen
  \bibfield  {author} {\bibinfo {author} {\bibfnamefont {D.~G.}\ \bibnamefont
  {Rothe}}, \bibinfo {author} {\bibfnamefont {R.~W.}\ \bibnamefont
  {Reinthaler}}, \bibinfo {author} {\bibfnamefont {C.-X.}\ \bibnamefont {Liu}},
  \bibinfo {author} {\bibfnamefont {L.~W.}\ \bibnamefont {Molenkamp}}, \bibinfo
  {author} {\bibfnamefont {S.-C.}\ \bibnamefont {Zhang}}, \ and\ \bibinfo
  {author} {\bibfnamefont {E.~M.}\ \bibnamefont {Hankiewicz}},\ }\href
  {http://stacks.iop.org/1367-2630/12/i=6/a=065012} {\bibfield  {journal}
  {\bibinfo  {journal} {New J. Phys.}\ }\textbf {\bibinfo {volume} {12}},\
  \bibinfo {pages} {065012} (\bibinfo {year} {2010})}\BibitemShut {NoStop}%
\bibitem [{\citenamefont {Tuttle}\ \emph {et~al.}(1989)\citenamefont {Tuttle},
  \citenamefont {Kroemer},\ and\ \citenamefont {English}}]{sm6}%
  \BibitemOpen
  \bibfield  {author} {\bibinfo {author} {\bibfnamefont {G.}~\bibnamefont
  {Tuttle}}, \bibinfo {author} {\bibfnamefont {H.}~\bibnamefont {Kroemer}}, \
  and\ \bibinfo {author} {\bibfnamefont {J.~H.}\ \bibnamefont {English}},\
  }\href {\doibase 10.1063/1.343167} {\bibfield  {journal} {\bibinfo  {journal}
  {J. Appl. Phys.}\ }\textbf {\bibinfo {volume} {65}},\ \bibinfo {pages} {5239}
  (\bibinfo {year} {1989})}\BibitemShut {NoStop}%
\bibitem [{\citenamefont {Gauer}\ \emph {et~al.}(1993)\citenamefont {Gauer},
  \citenamefont {Scriba}, \citenamefont {Wixforth}, \citenamefont {Kotthaus},
  \citenamefont {Nguyen}, \citenamefont {Tuttle}, \citenamefont {English},\
  and\ \citenamefont {Kroemer}}]{sm23}%
  \BibitemOpen
  \bibfield  {author} {\bibinfo {author} {\bibfnamefont {C.}~\bibnamefont
  {Gauer}}, \bibinfo {author} {\bibfnamefont {J.}~\bibnamefont {Scriba}},
  \bibinfo {author} {\bibfnamefont {A.}~\bibnamefont {Wixforth}}, \bibinfo
  {author} {\bibfnamefont {J.~P.}\ \bibnamefont {Kotthaus}}, \bibinfo {author}
  {\bibfnamefont {C.}~\bibnamefont {Nguyen}}, \bibinfo {author} {\bibfnamefont
  {G.}~\bibnamefont {Tuttle}}, \bibinfo {author} {\bibfnamefont {J.~H.}\
  \bibnamefont {English}}, \ and\ \bibinfo {author} {\bibfnamefont
  {H.}~\bibnamefont {Kroemer}},\ }\href {\doibase 10.1088/0268-1242/8/1S/031}
  {\bibfield  {journal} {\bibinfo  {journal} {Semicond. Sci. Technol.}\
  }\textbf {\bibinfo {volume} {8}},\ \bibinfo {pages} {S137} (\bibinfo {year}
  {1993})}\BibitemShut {NoStop}%
\bibitem [{\citenamefont {Aleshkin}\ \emph {et~al.}(2005)\citenamefont
  {Aleshkin}, \citenamefont {Gavrilenko}, \citenamefont {Gaponova},
  \citenamefont {Ikonnikov}, \citenamefont {Maremyanin}, \citenamefont
  {Morozov}, \citenamefont {Sadofyev}, \citenamefont {Johnson},\ and\
  \citenamefont {Zhang}}]{sm25}%
  \BibitemOpen
  \bibfield  {author} {\bibinfo {author} {\bibfnamefont {V.~Y.}\ \bibnamefont
  {Aleshkin}}, \bibinfo {author} {\bibfnamefont {V.~I.}\ \bibnamefont
  {Gavrilenko}}, \bibinfo {author} {\bibfnamefont {D.~M.}\ \bibnamefont
  {Gaponova}}, \bibinfo {author} {\bibfnamefont {A.~V.}\ \bibnamefont
  {Ikonnikov}}, \bibinfo {author} {\bibfnamefont {K.~V.}\ \bibnamefont
  {Maremyanin}}, \bibinfo {author} {\bibfnamefont {S.~V.}\ \bibnamefont
  {Morozov}}, \bibinfo {author} {\bibfnamefont {Y.~G.}\ \bibnamefont
  {Sadofyev}}, \bibinfo {author} {\bibfnamefont {S.~R.}\ \bibnamefont
  {Johnson}}, \ and\ \bibinfo {author} {\bibfnamefont {Y.~H.}\ \bibnamefont
  {Zhang}},\ }\href {\doibase 10.1134/1.1852637} {\bibfield  {journal}
  {\bibinfo  {journal} {Semiconductors}\ }\textbf {\bibinfo {volume} {39}},\
  \bibinfo {pages} {22} (\bibinfo {year} {2005})}\BibitemShut {NoStop}%
\bibitem [{\citenamefont {Gavrilenko}\ \emph {et~al.}(2010)\citenamefont
  {Gavrilenko}, \citenamefont {Ikonnikov}, \citenamefont {Krishtopenko},
  \citenamefont {Lastovkin}, \citenamefont {Maremyanin}, \citenamefont
  {Sadofyev},\ and\ \citenamefont {Spirin}}]{sm26}%
  \BibitemOpen
  \bibfield  {author} {\bibinfo {author} {\bibfnamefont {V.~I.}\ \bibnamefont
  {Gavrilenko}}, \bibinfo {author} {\bibfnamefont {A.~V.}\ \bibnamefont
  {Ikonnikov}}, \bibinfo {author} {\bibfnamefont {S.~S.}\ \bibnamefont
  {Krishtopenko}}, \bibinfo {author} {\bibfnamefont {A.~A.}\ \bibnamefont
  {Lastovkin}}, \bibinfo {author} {\bibfnamefont {K.~V.}\ \bibnamefont
  {Maremyanin}}, \bibinfo {author} {\bibfnamefont {Y.~G.}\ \bibnamefont
  {Sadofyev}}, \ and\ \bibinfo {author} {\bibfnamefont {K.~E.}\ \bibnamefont
  {Spirin}},\ }\href {\doibase 10.1134/S106378261005012X} {\bibfield  {journal}
  {\bibinfo  {journal} {Semiconductors}\ }\textbf {\bibinfo {volume} {44}},\
  \bibinfo {pages} {616} (\bibinfo {year} {2010})}\BibitemShut {NoStop}%
\bibitem [{\citenamefont {Tong}\ \emph {et~al.}(2017)\citenamefont {Tong},
  \citenamefont {Han}, \citenamefont {Li}, \citenamefont {Zhang}, \citenamefont
  {Sullivan},\ and\ \citenamefont {Du}}]{sm28}%
  \BibitemOpen
  \bibfield  {author} {\bibinfo {author} {\bibfnamefont {B.}~\bibnamefont
  {Tong}}, \bibinfo {author} {\bibfnamefont {Z.}~\bibnamefont {Han}}, \bibinfo
  {author} {\bibfnamefont {T.}~\bibnamefont {Li}}, \bibinfo {author}
  {\bibfnamefont {C.}~\bibnamefont {Zhang}}, \bibinfo {author} {\bibfnamefont
  {G.}~\bibnamefont {Sullivan}}, \ and\ \bibinfo {author} {\bibfnamefont
  {R.-R.}\ \bibnamefont {Du}},\ }\href {\doibase 10.1063/1.4993894} {\bibfield
  {journal} {\bibinfo  {journal} {AIP Advances}\ }\textbf {\bibinfo {volume}
  {7}},\ \bibinfo {pages} {075211} (\bibinfo {year} {2017})}\BibitemShut
  {NoStop}%
\bibitem [{\citenamefont {Knebl}\ \emph {et~al.}(2018)\citenamefont {Knebl},
  \citenamefont {Pfeffer}, \citenamefont {Schmid}, \citenamefont {Kamp},
  \citenamefont {Bastard}, \citenamefont {Batke}, \citenamefont {Worschech},
  \citenamefont {Hartmann},\ and\ \citenamefont {H\"ofling}}]{sm28a}%
  \BibitemOpen
  \bibfield  {author} {\bibinfo {author} {\bibfnamefont {G.}~\bibnamefont
  {Knebl}}, \bibinfo {author} {\bibfnamefont {P.}~\bibnamefont {Pfeffer}},
  \bibinfo {author} {\bibfnamefont {S.}~\bibnamefont {Schmid}}, \bibinfo
  {author} {\bibfnamefont {M.}~\bibnamefont {Kamp}}, \bibinfo {author}
  {\bibfnamefont {G.}~\bibnamefont {Bastard}}, \bibinfo {author} {\bibfnamefont
  {E.}~\bibnamefont {Batke}}, \bibinfo {author} {\bibfnamefont
  {L.}~\bibnamefont {Worschech}}, \bibinfo {author} {\bibfnamefont
  {F.}~\bibnamefont {Hartmann}}, \ and\ \bibinfo {author} {\bibfnamefont
  {S.}~\bibnamefont {H\"ofling}},\ }\href {\doibase 10.1103/PhysRevB.98.041301}
  {\bibfield  {journal} {\bibinfo  {journal} {Phys. Rev. B}\ }\textbf {\bibinfo
  {volume} {98}},\ \bibinfo {pages} {041301} (\bibinfo {year}
  {2018})}\BibitemShut {NoStop}%
\end{thebibliography}

%

\end{document}